\documentclass[apjl,twocolumn,twocolappendix,showkeys,numberedappendix]{oja} %for oja

\usepackage{newtxtext, newtxmath}
\usepackage[T1]{fontenc} 

\DeclareRobustCommand{\VAN}[3]{#2}
\let\VANthebibliography\thebibliography
\def\thebibliography{\DeclareRobustCommand{\VAN}[3]{##3}\VANthebibliography}

\usepackage{comment}
\usepackage{bm}
\usepackage {amsmath} \usepackage{graphicx}	  
\usepackage{sidecap}   
\usepackage{caption}   \usepackage{subcaption}   %\usepackage{fontspec}  \usepackage{setspace}
\usepackage{ragged2e} 
\usepackage[breaklinks,colorlinks,citecolor=blue,urlcolor=blue]{hyperref}
\usepackage{float} 

\usepackage{graphicx}
\graphicspath{./figures/}
\newcommand\blfootnote[1]{%
	\begingroup
	\renewcommand\thefootnote{}\footnote{#1}%
	\addtocounter{footnote}{-1}%
	\endgroup
}
\newcommand{\nnn}{\hat{\mbox{\boldmath $n$}} {}}
\newcommand{\xx}{\bm{x}}
\def\Brms{B_{\rm rms}}
\newcommand{\rr}{\mbox{\boldmath $r$} {}}
\newcommand{\EQ}{\begin{equation}}
\newcommand{\EN}{\end{equation}}
\newcommand{\EQA}{\begin{eqnarray}}
\newcommand{\ENA}{\end{eqnarray}}

\newcommand{\Eq}[1]{Eq.~(\ref{#1})}
\newcommand{\Eqs}[2]{Eqs.~(\ref{#1}) and~(\ref{#2})}
\newcommand{\Fig}[1]{Fig.~\ref{#1}}

\newcommand{\Figs}[2]{Figs.~\ref{#1} and \ref{#2}}

\newcommand{\Tab}[1]{Table~\ref{#1}}
\newcommand{\bra}[1]{\langle #1\rangle}

\newcommand{\BB}{\bm{B}}

\newcommand{\uu}{\bm{u}}

\newcommand{\AAA}{\bm{A}}
\newcommand{\JJ}{\bm{J}}
\newcommand{\nab}{\bm{\nabla}}
\begin{document}
	\title{Quasi-two-dimensionality of three-dimensional, magnetically dominated, decaying turbulence \vspace{-1.5cm}}
	\author{Shreya Dwivedi$^{1,2,\dag}$}
	\author{Chandranathan Anandavijayan$^{1,\star}$}
	\author{Pallavi Bhat$^{1,\ddag}$}
		
		% List of institutions
		\affiliation{$^{1}$International Centre for Theoretical Sciences, Tata Institute of Fundamental Research, Bangalore 560089, India}
		\affiliation{$^{2}$Department of Physics, University of Wisconsin-Madison, Madison, WI 53706, USA
	}
	% These dates will be filled out by the publisher
	%\date{Accepted XXX. Received YYY; in original form ZZZ}
	
	% Enter the current year, for the copyright statements etc.
	%\pubyear{2024}
	
	%% Don't change these lines
	%\begin{document}
	
	%\maketitle
	%\label{firstpage}
	%\pagerange{\pageref{firstpage}--\pageref{lastpage}}
	% Abstract of the paper
\begin{abstract}
Decaying magnetohydrodynamic (MHD) turbulence is important in various astrophysical contexts, including early universe magnetic fields, star formation, turbulence in galaxy clusters, magnetospheres and solar corona. Previously known in the nonhelical case of magnetically dominated decaying turbulence, we show that magnetic reconnection is important also in the fully helical case and is likely the agent responsible for the inverse transfer of energy. Again, in the fully helical case, we find that there is a similarity in power law decay exponents in both 2.5D and 3D simulations. To understand this intriguing similarity, we investigate the possible quasi-two-dimensionalization of the 3D system. We perform Minkowski functional analysis and find that the characteristic length scales of a typical magnetic structure in the system are widely different, suggesting the existence of local anisotropies. Finally, we provide a quasi-two-dimensional hierarchical merger model which recovers the relevant power law scalings. In the nonhelical case, we show that a helicity-based invariant cannot constrain the system, and the best candidate is still anastrophy or vector potential squared, which is consistent with the quasi-two-dimensionalization of the system.
	\end{abstract}
	% Select between one and six entries from the list of approved keywords.
	% Don't make up new ones.
	
	%---------------------------------------------------------------------------------------------------------------------------------------------
	\blfootnote{\hspace{-0.5 cm} $\dag$ \href{mailto:sdwivedi4@wisc.edu}{sdwivedi4@wisc.edu} \\
		$\star$ \href{mailto:chandranathan.a@icts.res.in}{chandranathan.a@icts.res.in}
		\\
		$\ddag$ \href{mailto:pallavi.bhat@icts.res.in}{pallavi.bhat@icts.res.in}
	}
\maketitle	
%------------------------------ 
\section{Introduction}
%[decaying turbulence arises in many sitchs. Relaxe via reconnection. What is reconnection]
Astrophysical plasmas are typically in a state of turbulence. 
In the aftermath of energetic processes that are responsible for generation of the turbulence, the 
plasma undergoes relaxation, leading to the decay of velocity and magnetic fields. 
Many astrophysical plasmas such as solar and stellar coronae, accretion disks, and pulsar 
magnetospheres host magnetic fields of 
significant strength and are magnetically dominated \citep{pevtsov_relationship_2003, blandford_magnetoluminescence_2017, narayan_magnetically_2003}.
%solar corona, accretion disks, pulsar magnetospheres, magnetars, 
Such magnetic turbulent systems are thought to relax via magnetic reconnection \citep{taylor_relaxation_1986}. 
Reconnection is a process by which magnetic fields undergo topological 
reconfiguration accompanied by conversion of magnetic energy into heat, radiation and 
particle acceleration. These can be highly energetic events and are thought to  
explain solar flares, magnetospheric substorms, gamma-ray bursts etc \citep{ruan_fully_2020, hesse_magnetic_2020, mckinney_reconnection_2012}. 

%[timescale reconnection. Relaxation. What are the sitches for decaying turb]
More recently, it was shown that magnetic reconnection plays an important role 
in 3D MHD decaying turbulence \citep{bhat_inverse_2021, zhou_multi-scale_2020}
further extending support 
to the idea of reconnection playing an important role in relaxation.
Some of the astrophysical 
scenarios are either theoretically modelled or already well understood to 
relax via magnetic reconnections.
%solar
For example, in the Sun, magnetic fields in the form of flux ropes extending up into the corona 
experience twists due to the motion of the foot points in the photosphere. 
The Parker model of coronal heating directly employs 
magnetic reconnection to relax such twisted magnetic flux ropes \citep{parker_magnetic_1983}. 
Inevitably, the nonlinear evolution of reconnecting flux ropes leads to turbulence \citep{rappazzo_nonlinear_2008}, 
accompanied by growth in magnetic structure size \citep{pontin_dynamics_2011}. 
%earth magnetosphere
Another context is the interaction between solar wind and earth's magnetotail, which leads to reconnection.
This involves magnetic flux loading and subsequent release of energy resulting in magnetospheric substorms \citep{mcpherron_magnetospheric_1979}. 
Thereafter, the earth's magnetic field relaxes to its ground state, and these processes again 
feature turbulence \citep{daughton_role_2011}.
%pulsar
In pulsar wind nebulae, the striped wind and large magnetic hoops are thought to 
relax via reconnections in decaying turbulence, leading to nonthermal emission and flares \citep{zrake_crab_2016, zrake_turbulent_2017}. 
Additionally, the magnetic dissipation due to post-shock instabilities seems to explain the lower strength of the magnetic field from  
observations \citep{porth_three-dimensional_2014}. 

%[inverse transfer. Selective decay]
Magnetic reconnection, as a model for relaxation, can also naturally explain 
the phenomenon of inverse energy transfer. Inverse energy transfer occurs when energy is transferred to larger scales as the turbulent system relaxes. In the 3D hydrodynamic case, relaxation of homogeneous and isotropic turbulence occurs by selective decay, i.e. energy at the smallest scales decays first, followed by decay of successively larger scales. This process does not lead to any inverse energy transfer but depicts what is known as the “permanence of large eddies” \citep{batchelor_role_1949, LL1975, lesieur_3d_2000}. On the other hand, it is well known that decaying 2D hydrodynamic turbulence can depict inverse energy transfer because here a different version of selective decay manifests. Here, one ideal invariant (kinetic energy) decays more slowly than another (enstrophy). In fact, the slower decay invariably is due to its condensation at larger scales, thus being consistent with the original understanding of selective decay \citep{matthaeus_selective_1980, ting_turbulent_1986}. While permanence of large eddies just constitutes the continuation of the flow at larger scales (or smaller wavenumbers that are a part of the forward slope) at similar amplitudes, in other cases like 2D hydrodynamics or 3D MHD with magnetic helicity, the slower decaying ideal invariant can cause an actual inverse transfer in energy (involving increase of energy at larger scales or smaller wavenumbers) \citep{mininni_inverse_2013}. 

%[2D it is clear reconnection. But also 3D. Coherence increase useful to understand some astro]
In the MHD case, it is easy to observe that in a 2D decaying system, an inverse transfer emerges as a consequence of reconnection between magnetic islands \citep{zhou_magnetic_2019}. 
A similar process in 3D could lead to larger and larger magnetic structures at the cost of 
loss/decay in magnetic energy. The study in \citet{bhat_inverse_2021} (BZL21)
shed light on the possibility of magnetic reconnection driving such a 3D inverse energy transfer. 
BZL21 were motivated by the observation that the decay laws in 3D simulations matched with that 
seen in 2D. They identified the dynamical timescale in this system of nonhelical decaying turbulence as the reconnection timescale, $\tau_{rec}$. They found that the evolution curves collapsed when the time was normalized by the reconnection timescale. 
This result rules out the conventional notion that the dynamical time scale in a magnetically 
dominated system is typically the Alfv\'enic time scale. Thus, the inverse transfer phenomenon was shown to occur on a slower time scale. 

The inverse transfer phenomenon is useful in astrophysical 
scenarios where the coherence length scale of the magnetic field is important as well. 
The ISM magnetic fields that first feed onto a central massive object need to be sufficiently coherent to trigger magnetorotational instability needed for angular momentum transport in accretion disks \citep{bhat_evolution_2017}. In galaxy clusters, large Faraday rotation measures need coherent magnetic fields \citep{subramanian_evolving_2006, bhat_fluctuation_2013}. In IGM, the non-detection of GeV photons that should have originated from cascading of TeV photons emanating from blazars again needs coherent magnetic fields for deflecting charged particles \citep{neronov_evidence_2010, subramanian_origin_2016, hosking_cosmic-void_2023}. The magnetic fields in ISM, galaxy clusters and IGM, as mentioned in the examples above, could have benefited from an inverse transfer in magnetic energy during a turbulent decaying phase (when the dynamo was inactive). 

%[what recent results show exactly. 2D-3D match. H&S]
Direct numerical simulations of magnetically dominated turbulence have shown evidence for inverse 
energy transfer even when an ideally conserved quantity like magnetic helicity was absent \citep{brandenburg_nonhelical_2015, zrake_inverse_2014}. BZL21 postulated that anastrophy conservation (in the absence of magnetic helicity) is responsible for the inverse transfer phenomenon in 3D. They highlighted the link between anastrophy conservation in the large Lundquist number limit and 
magnetic reconnection in 2D and claimed that this carries over to the case of 3D as well, 
thus explaining the exact similarity in the decay law between 2D and 3D. 
However, \citet{hosking_reconnection-controlled_2021} introduced a new integral they call the Saffman helicity invariant based on magnetic helicity fluctuations, which they claimed is responsible for constraining the decay law (this does not give $t^{-1}$, but gives $t^{-1.18}$ instead). 
They extended the BZL21 result of $\tau_{rec}$ (reconnection timescale) being the dynamical timescale to the helical case as well 
and found that it leads to a $t^{-4/7}$ decay law. 

In our work here, we examine helical decaying MHD turbulence and find that, indeed, reconnection 
is important in this case as well. It makes sense that reconnection is a common 
explanation for the phenomenon of structures growing larger and larger as the decay progresses 
in both helical and nonhelical inverse energy transfer. 
Intriguingly, the 2D(2.5D)-3D correspondence holds in the helical case as well; 
the decay law in 2.5D (magnetic helicity cannot be defined in the strict 2D case) is the same as that seen in 3D. We also assess the so-called Saffman helicity integral critically and find that it is most likely ruled out given the 2D-3D correspondence. 

To assess the quasi-two-dimensional nature of the magnetically dominated decaying turbulence, we analyse the system using Minkowski functionals. These were originally introduced as a tool to study spatial patterns in galaxy cluster data \citep{mecke_robust_1994}. In cosmology, it is well-known that there is clustering of galaxies \citep{Peebles1980}. It was found that the two-point correlation function for large-scale galaxy clustering is insufficient due to non-gaussianity. Minkowski functionals, on the other hand, contain information related to all the correlation functions of the order and above two-point correlation \citep{wiegand_direct_2014}. In this paper, we use these Minkowski functionals to examine the morphology of reconnecting fields and the associated current density structures. This tool was first used in this manner of assessing shapes in a scalar field in the field of cosmology and structure formation \citep{sahni_shapefinders_1998}. It has also been previously used in the field of astrophysical fluid dynamics for analysing magnetic field structures in simulations of small-scale dynamo \citep{wilkin_magnetic_2007, seta_saturation_2020}. And to assess current density structures in strong MHD turbulence \citep{zhdankin_energy_2014}. 
We primarily employ Minkowski functionals to quantify the inherent anisotropy within the system. This approach aids in refining our comprehension of the quasi-two-dimensional manifestation of the reconnections. 

The following section covers the MHD equations, initial conditions and parameters employed in our direct numerical simulations. Section 3 delves into the analysis of data and results, with subsection 3.1 dedicated to the helical case and subsection 3.2 focusing on the nonhelical case. Within these subsections, we present evidence supporting the 2D-3D similarity in decay laws. Subsequent analyses compare conserved quantities and critically evaluate various integrals believed to constrain the evolution of these decaying systems. Section 4 presents results from the Minkowski functional analysis, followed by the introduction of an analytical quasi-2D hierarchical merger model in Section 5. The concluding sections, 6 and 7, consist of discussions and overall conclusions respectively.
In the Appendix Sections A and B, we delineate the algorithms employed for Minkowski functional analysis. 

\section{Numerical Setup}

\subsection{The model}
We have run direct numerical simulations of decaying helical and nonhelical MHD turbulence in 2.5D and 3D cubes, using \textsc{Pencil-Code}\footnote{http://pencil-code.nordita.org}, with periodic boundary conditions. We solve the continuity equation,  momentum equation and the uncurled version of induction equations on a Cartesian $N^2$ or $N^3$ grid, where $N$ is the number of grid points in any direction.  
\begin{eqnarray}
\frac{D~\mathrm{ln}~\rho}{Dt} &=& - \nab \cdot \uu, \\
\frac{D\uu}{Dt} &=& -c_s^2\nab~\mathrm{ln}~\rho + \frac{\JJ\times \BB}{\rho} + \frac{\bm{F}_{visc}}{\rho},\\
\frac{\partial\AAA}{\partial t} &=& \uu \times \BB - \eta \mu_0 \JJ,
\end{eqnarray}
where $D/Dt = \partial/\partial t + \uu\cdot\nab$ is the advective derivative, $\uu$ is the fluid velocity field, $\BB = \nab \times \AAA$ is the magnetic field, with $\AAA$ the magnetic vector potential, and $\JJ = \nabla \times \BB/\mu_0$ is the current density, with $\mu_0$ the vacuum permeability, $\rho$ is the mass density, and $\eta$ is the magnetic diffusivity. The scalar potential $\Phi = 0$ (Weyl gauge), and the viscous force $\bm{F}_{visc} = \nab \cdot 2\nu \rho \bm{S}$, where $\nu $ is the kinematic viscosity, and $\bm{S}$ is traceless rate of strain tensor with components $\bm{S}_{ij} = \frac{1}{2} (u_{i,j} + u_{j,i}) - \frac{1}{3} \delta_{ij}\nab \cdot \uu$ (commas denote partial derivatives).

 A 2.5 simulation is similar to 2D, except the evolving characteristic field vectors have all three components along $x$, $y$ and $z$. For instance, $\BB = \nabla \times \AAA(x, y) = [B_x (x, y), \, B_y (x, y), \, B_z (x, y)]$.

While the system is isothermal with adiabatic index $\gamma = 1$, the variables $\nu$, $\eta$ and mean $\rho$ remain constant throughout the decaying turbulence. All the quantities obtained from a simulation are dimensionless, characterized by length, velocity, density, and magnetic field, measured in the following units: system size ($L$), isothermal sound speed ($c_s$), initial density ($\rho_0$), and $\sqrt{\mu_0 \rho_0 c_s^2}$ respectively. Here, $L = 2\pi$ and $\rho_0 = \mu_0 = c_s = 1$.

\begin{table}
\begin{center}
\begin{tabular}{ ||c|c|c|c|c|c|c|| } 
 \hline \hline
 Helicity & Resolution & $\eta \times 10^{-4}$ & $\nu \times 10^{-4}$ &$u_{rms_0}$ & $B_{rms_0}$ & $S/10^3$\\ [0.2ex]
 \hline 
 nonhel & $1024^2$ & 1.3 & 1.3 & 0. & 0.4 & 0.8\\ 
 nonhel & $2048^2$ & 0.34 & 0.34 & 0. & 0.4 & 3.\\
 nonhel & $4096^2$ & 0.168 & 0.168 & 0. & 0.4 & 6.\\
 hel & $1024^2$ & 1.3 & 1.3 & 0. & 0.4 & 0.8\\ 
 hel & $2048^2$ & 0.34 & 0.34 & 0. & 0.4 & 3.\\
 hel & $4096^2$ & 0.168 & 0.168 & 0. & 0.4 & 6.\\
 nonhel & $512^3$  & 5.4 & 5.4 & 0. & 0.4 & 0.2\\
 nonhel & $1024^3$ & 2.5 & 2.5 & 0. & 0.4 & 0.4\\
 nonhel & $1024^3$ & 1.3 & 1.3 & 0. & 0.4 & 0.8\\
 hel & $512^3$ & 5.4 & 5.4 & 0. & 0.4 & 0.2\\
 hel & $1024^3$ & 2.5 & 2.5 & 0. & 0.4 & 0.4\\
 hel & $1024^3$ & 1.3 & 1.3 & 0. & 0.4 & 0.8\\ [0.5ex] 
 \hline 
 \hline
\end{tabular}
\end{center}
\caption{A summary of all the runs is listed. Helicity refers to whether the initial random magnetic field is fully helical (given by 'hel') or nonhelical (given by 'nonhel').}
\label{tab1}
\end{table}

\subsection{Initial conditions and parameters}
\label{initSim}
We initialize the magnetic vector potential component along $j$ in the Fourier domain as follows,
\\
\\
Nonhelical magnetic case:
\EQ
    \hat{A}_j (\mathbf{k}) = h_j(\mathbf{k}) (k/k_{cf})^{n/2 -1}  \\  
    \label{aj_nonhel}
\EN
Helical magnetic case:
%\EQ
%    \hat{A}_j (\mathbf{k}) = \left[\mathbf{P}_{jl} - i\sigma_M\epsilon_{jlm} \frac{k_m}{k}\right] \frac{k_p^{-3/2} h_j(\mathbf{k}) (k/k_p)^{n/2 -1}} {[1 + (k^2/k_p^2)^{(n+11/3)}]^{1/4} } \\  
%    \label{aj_hel}
%\EN
\EQ
    \hat{A}_j (\mathbf{k}) = \left[\mathbf{P}_{jl} - i\sigma_M\epsilon_{jlm} \frac{k_m}{k}\right] h_j(\mathbf{k}) (k/k_{cf})^{n/2 -1} \\  
    \label{aj_hel}
\EN
\\
where $j \in \{x, y, z\}$, $\mathbf{P}_{jl} = \delta_{jl} - k_j k_l/k^2$ is the projection operator,  $\sigma_M = 1$ is the relative magnetic helicity, inducing fully helical magnetic field at $t=0$, and
\EQ
    h_j(\mathbf{k}) = A_0 \exp{(i\phi(\mathbf{k}))} \exp{\left(-\frac{1}{2} \frac{k^2}{k_{cf}^2}\right)} \\
    \label{h_j}
\EN
is a spectrum with amplitude $A_0$, random phases $\phi \mathbf{(k)}$ (between -$\pi$ to $\pi$) and Gaussian distributed fluctuations that exponentially cuts off beyond $k_{cf}$. This generates the shell integrated 1D magnetic power spectrum $M(k)$ that grows with a slope $k^\alpha$, where the exponent $\alpha=4$, and peaks at $k_p \sim \frac{1}{L_{int}} \simeq 25$, over small scales at $t = 0$, where $L_{int}$ is the integral length scale calculated as,
\EQ
L_{int} = \frac {\int (2\pi / k) M(k) dk } {\int M(k) dk},
\label{lint}
\EN
and the initial root mean squared magnetic field $B_{rms_0} = [2\int M(k, t=0) dk]^{1/2} \simeq 0.4$, the initial velocity field $u_{rms_0} = 0$, and the magnetic Prandtl number $=1$ for all the helical and nonhelical runs referred to in this paper.

The Lundquist number $S=(2\pi k_p^{-1})\sqrt{2E_M}/\eta $, where $E_M$ is total magnetic energy. The initial parameters and configuration of the numerical simulations run on $1024^2, 2048^2$ and $4096^2$ grids in 2.5D, and $512^3$ and $1024^3$ grids in 3D, with high $S$, are shown in \Tab{tab1}.

With the evolution of time, $M(k, t)$ decays with a slope of $k^{-5/2}$ for the helical magnetic decaying turbulence\footnote{refer \Fig{ps_hel} for power spectrum at different time snapshots}, whereas with a slope of $k^{-2}$ for the nonhelical case (shown in a previous paper by \citet{bhat_inverse_2021} (BZL21). 

\section{Results}
\subsection{Reconnection in helical decaying MHD turbulence}
\label{helicalcase}

\begin{figure}%[h!]
\begin{flushright}
\includegraphics{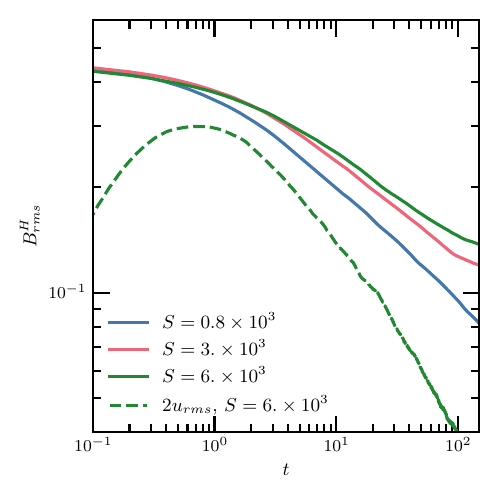} \\
\includegraphics{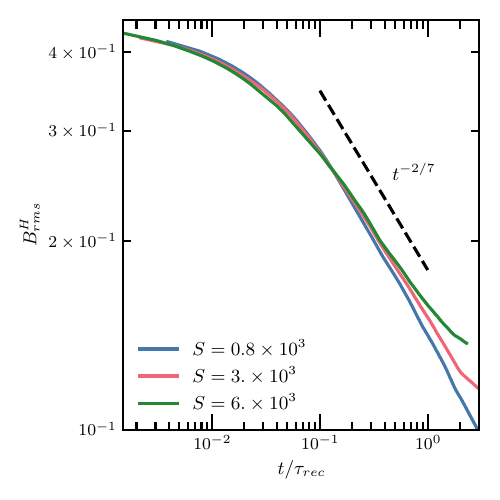}
\end{flushright}
%\end{minipage}
\caption{Upper panel: evolution curves for $B^H_{rms}$ and $u_{rms}$ vs t from fully helical 2.5D runs with different values for $S$. Lower panel: collapse of the $B^H_{rms}$ evolution curves when plotted against $t/\tau_{rec}$.}  
\label{2.5dplots}
\end{figure}

\begin{figure}
\begin{flushright}
\includegraphics{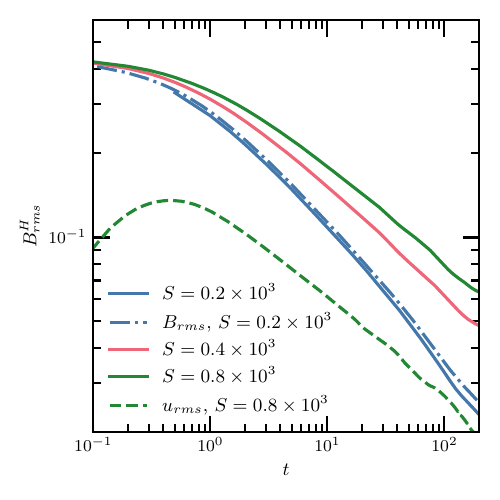}
\includegraphics{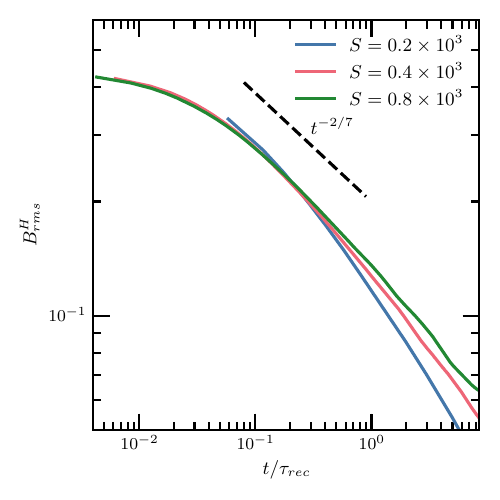}
\caption{Upper panel: evolution curves for $B^H_{rms}$ and $u_{rms}$ vs t from fully helical 3D runs with different values for $S$. Lower panel: collapse of the $B^H_{rms}$ evolution curves when plotted against $t/\tau_{rec}$.}
\label{3dplots}
\end{flushright}
\end{figure}

It has been well known that in a decaying turbulent system, if the magnetic fields are helical, 
then they tend to relax to larger scales. This results in an inverse cascade of 
magnetic helicity accompanied by a simultaneous inverse transfer of magnetic energy \citep{pouquet_frisch_léorat_1976}. 
The physical picture typically discussed involves twists or writhes in the field, 
relaxing to larger scales over a period of time.  In this context, motivated by our 
previous results in BZL21, we explore if magnetic reconnection has a role 
to play in helical decaying MHD turbulence as well. 

In \Figs{2.5dplots}{3dplots}, we show results from both 2.5D and 3D simulations. These were 
initialized with zero velocity field\footnote{refer to \Fig{urms_plots} in the Appendix to view zero initial velocity field.} and random fully helical magnetic fields peaked at small scales which then decays as the simulations progress.  In particular, we show the evolution curves of the rms value of the helical part of the magnetic field, defined as, 
\EQ
B^H_{\rm rms} = \sqrt {2 E_{M}^H}, \quad E_{M}^H =  \frac{1}{2} \int k H(k) dk,
\EN
where $H(k)$ is the shell-averaged 1D spectrum of the magnetic helicity. 
In both the plots for 2.5D and 3D cases in \Figs{2.5dplots}{3dplots}, the difference between the top and bottom panels is that 
in the latter, the time axis is normalized by the magnetic reconnection time-scale $\tau_{rec}=S^{1/2}\tau_{A}$, 
where $\tau_{A}=(2\pi k_p^{-1})/\sqrt{2E_M})$ is the Alfv\'en time-scale. We observe that the $B^H_{\rm rms}$ evolution curves 
from runs with different Lundquist numbers collapse on top of each other. This reveals that the relevant dynamical time-scale is indeed the reconnection time-scale. 
This kind of collapse was first seen in Figs. 10 and 11 in BZL21, in the nonhelical decaying MHD turbulence case. 
Although the collapse in 3D is not as smooth as that in the 2.5D case, 
it can be inferred that the curves would asymptotically collapse at significantly higher Lundquist numbers. 
MHD simulations in 3D at Lundquist numbers larger than $800$, i.e. at resolutions higher than $1024^3$, could not be conducted because of HPC resource constraints.
The collapsing curves form the first part of the evidence for reconnection 
playing a role in decaying helical MHD turbulence. Next we study the slope of the 
decay curve. 

One can calculate how the magnetic energy scales with time using simple considerations 
of power law decay (this is a standard method in hydrodynamics to derive decay scalings \citep{kolmogorov_dissipation_1997}), 
\EQ
\frac{dE_M}{dt} \sim - \frac{E_M}{\tau_{\rm rec}}
\label{decaylaw}
\EN
and the conservation of magnetic helicity, $\int \AAA \cdot \BB ~dV$. 
Typically, the system under consideration is initialized to have 
a spectrum that rises as $k^4$, peaks at large wavenumbers, and falls quickly. 
Such a system, as it decays, retains this type of spectral shape with the peak shifting to the smaller 
wavenumbers. This indicates that the system, at any given time, has statistically more structures 
pertaining to the integral length scale or scales close to that corresponding to the peak wavenumber, $k_p$. This allows us to simplify and approximate our mathematical descriptions to 
follow the dynamics of the fields concentrated at $k_p$. With this in mind, one can simplify the conservation 
of magnetic helicity as, 
\EQ
E_M k_p^{-1} \sim const.
\label{maghelcons}
\EN
Thus, with \Eqs{decaylaw}{maghelcons}, we obtain $E_M \sim t^{-4/7}$ scaling. 
This scaling using $\tau_{rec}$ was recently also derived 
by \citet{hosking_reconnection-controlled_2021}. 
We find that this scaling bears out in both 2.5D and 3D simulations as is seen in \Figs{2.5dplots}{3dplots}. This further strengthens the idea that magnetic reconnection drives the decay and leads to the inverse 
transfer of energy in this system. 

\begin{figure}
\begin{flushright}
\includegraphics{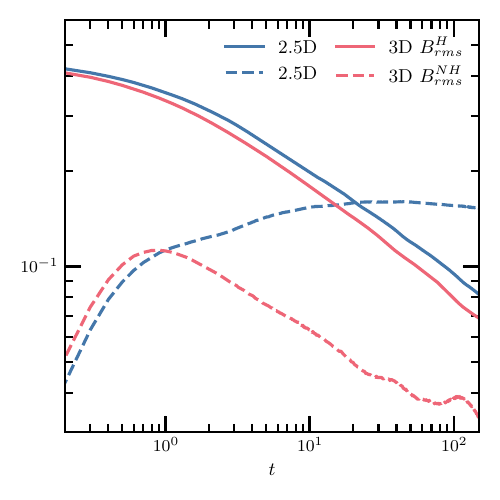}
\end{flushright}
\caption{Comparison of evolution curves for $B^H_{rms}$ from helical 2.5D and 3D runs at the same value of $S=800$.}  
\label{label-c}
\end{figure}

%about the matching results between 2D and 3D
An important point deserving of attention is that the scaling of magnetic energy with time is the same in both 2.5D and 3D, a fact that has not been adequately acknowledged. This cannot be a coincidence, given that such a similarity 
was also found in the case of nonhelical decaying MHD turbulence in BZL21. 
In \Fig{label-c}, we show the evolution curves from both 2.5D and 3D simulations together to clearly see that the slopes match. 
The underpinnings of this similarity between 2.5D and 3D will be discussed in detail in a later
section ~\ref{localaniso}. 

%\begin{minipage}
\begin{figure}
\begin{flushright}
\includegraphics{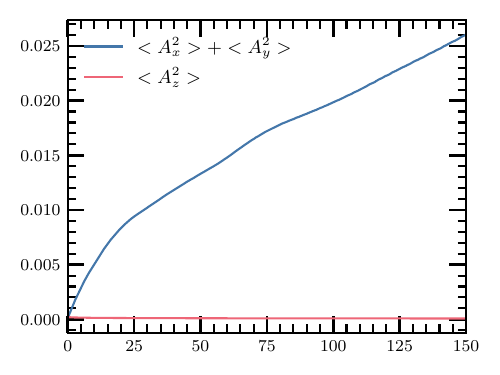}\\
\includegraphics{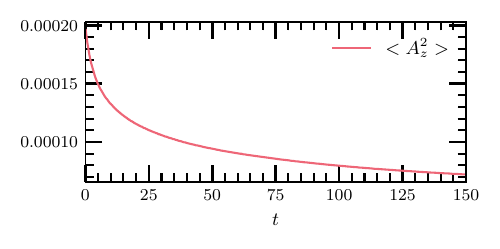}
\end{flushright}
\caption{Evolution of $\bra{A_x^2$ + $A_y^2}$ and $\bra{A_z^2}$ from 2.5D helical run at ${1024}^2$ resolution with $S=800$.}  
\label{label-d}
\end{figure}

\begin{figure*}[t]
\centering
\includegraphics[width = 0.32\textwidth]{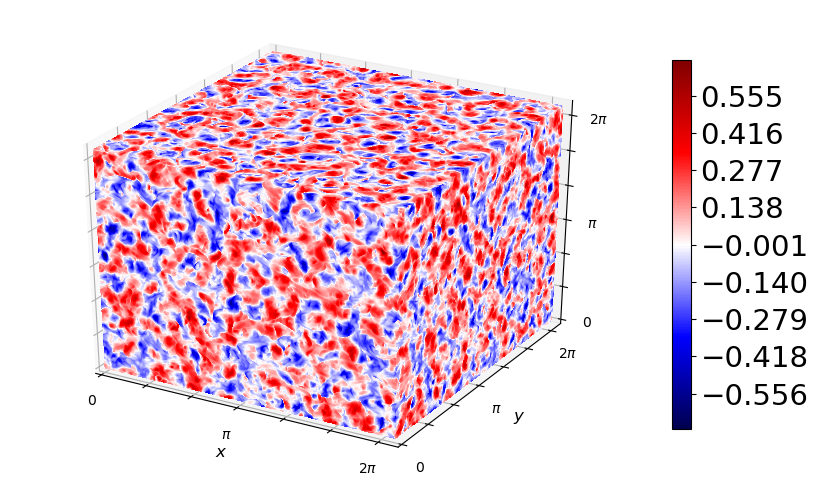}
\includegraphics[width = 0.32\textwidth]{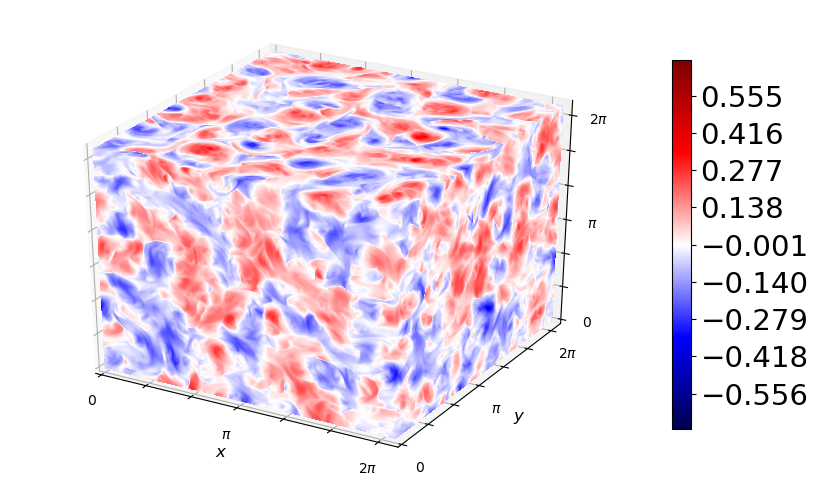}
\includegraphics[width = 0.32\textwidth]{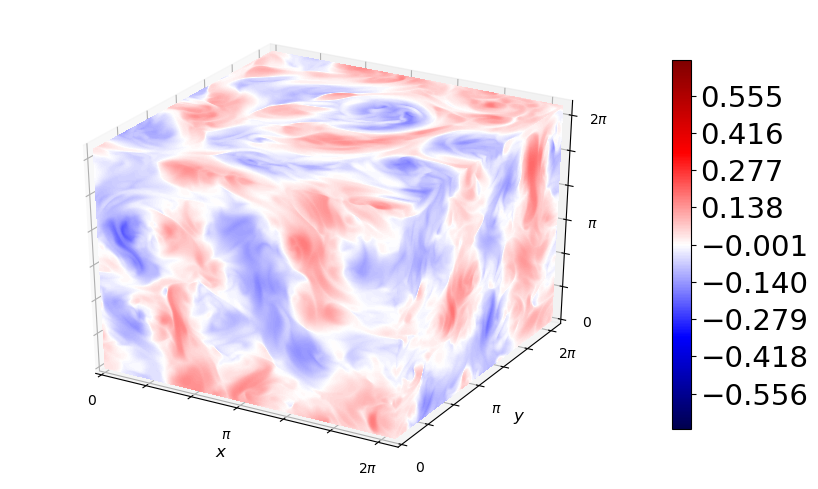}
\caption{Evolution of $B_z$ on 3D domain from 3D helical ($1024^3$, $S=800$) run at times $t = 2, 12$ and $46$ from left to right.}
\label{boxevol}
\end{figure*}

The reason we explicitly calculated the helical part of the $B_{\rm rms}$ is that 
in the 2.5D case, the nonhelical part has a nontrivial evolution curve and thus total $\Brms$ 
is not dominantly helical throughout the simulation. 
In \Fig{label-c}, we show the evolution curves (dashed lines) for the nonhelical part of the field, $B_{\rm rms}^{NH} = \sqrt {2 E_{M}^{NH}}$
computed by subtracting out the helical part of the energy from the total, $E_{M}^{NH} = E_M-E_{M}^H$. 
It can be seen that, in the 3D case, after an initial rise around $t\sim 1$, $B_{\rm rms}^{NH}$ gradually decreases until the end of the simulation. In contrast, in the 2D case, $B_{\rm rms}^{NH}$ shows growth over time. 
This growth is because even though vector-potential squared, 
$ \int \bra{A_z^2} dV$ is a conserved quantity in 2D, 
since it is actually a 2.5D simulation (and not strictly 2D), the other two components 
of the vector potential corresponding to $B_z$ are allowed to grow as is shown in \Fig{label-d}. 
However, from anti-dynamo theorems, we expect this growth to cease and $B_z$ should also decay ultimately \citep{kandubook}.

\begin{figure}[h!]
    \begin{flushright}
    \includegraphics{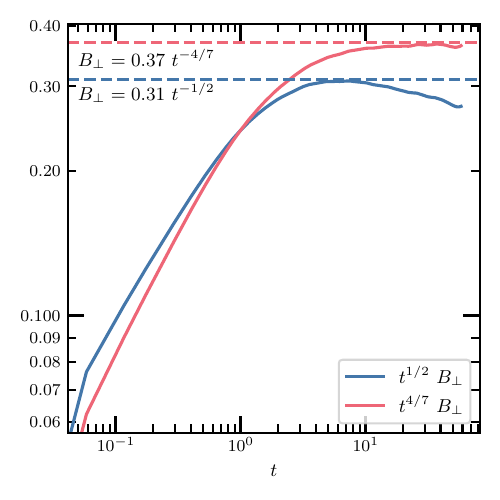}
    \includegraphics{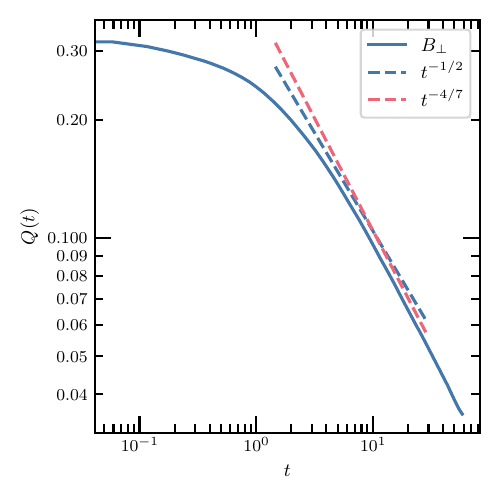}
    \end{flushright}
    \caption{Evolution curve for perpendicular component of $\bm{B}$ for a $4096^2$ helical run with $S=800$, showing that $B_{\perp}$ decays as $t^{-4/7}$.}
    \label{fig:hel 2.5D bz}
\end{figure}

At this point, it is important to observe that if we were to write down the 
MHD equations separately for the component of the fields which are in-plane separately (i.e. fields that are perpendicular to $z$ and thus can be denoted as $\BB_{\perp}$), they exactly match with equations for a strictly 2D case and thus 
one expects from anastrophy conservation due to $\bra{A_z^2}$ (also seen from \Fig{label-d}), a scaling law, 
\EQ
B_\perp k_p^{-1} \sim const.
\label{bperp}
\EN 
While this is true, only $B_\perp$ component evolution is constrained by \Eq{bperp} and ultimately, it is the magnetic helicity which will constrain the evolution of $B_z$ and $B_\perp$ simultaneously and 
thus the primary scaling law is given by, 
\EQ
B_{\rm z} B_\perp k_p^{-1} = B_{\rm rms}^{H} k_p^{-1} \sim const.
\label{maghelscaling}
\EN 
The implications of the two \Eqs{bperp}{maghelscaling} are that $B_z  \sim const.$ and $B_\perp \sim t^{-4/7}$. If, instead, one were to take \Eq{bperp} to be 
the primary scaling law, one would obtain $B_\perp \sim t^{-1/2}$. We confirm in \Fig{fig:hel 2.5D bz}, that simulation data verifies the decay exponents of $-4/7$ and $-2/7$ for 
$B_\perp$ and $B_{\rm rms}^{H}$ respectively instead of $-1/2$ and $-1/4$. 

Nonetheless, it is interesting that in the 2.5D case, nonhelical energy is produced in a system trying to achieve relaxation via 
reconnection. So, in fact, the total field decays at a much slower rate. 
In the 3D case, nonhelical energy is not produced as much. Thus, the decay of the total field reflects mainly the evolution of the helical component, 
which, as we have mentioned earlier, has the same scaling in time as seen in the 2.5D case. 

In the 2.5D case, the contour plots of the vector potential $A_z$ strung into a movie
\footnote{The link to 2.5D helical $4096^2$ simulation movie: \href{https://youtu.be/AyUrKS4kl-o}{$A_z$ contour plots.}} 
show the merging of the magnetic islands with time. Localized current sheets form between 
two merging islands, leading to magnetic reconnection and, thus, larger islands. Similarly, plots of magnetic structures on a 3D domain at a range of time instances for each magnetic field component $[B_x, B_y, B_z]$ from helical 3D simulations are strung into a movie
    \footnote{The links to 3D helical $1024^3$ sim movie: 
    \href{https://youtu.be/XWSgYKJ6L80}{$B_x$}
    \href{https://youtu.be/LN4YIanJuNM}{$B_y$} 
    \href{https://youtu.be/jze4wUadt3g}{$B_z$}.}. 
    This movie shows initial small-scale 
magnetic structures, moving with the turbulent flow merging with 
the localized neighbouring like-signed structures, thereby forming larger structures with the evolution of time as seen in \Fig{boxevol}. 
The structures merge likely via magnetic reconnection, leading to an inverse transfer of magnetic energy.   

\subsection{The case of nonhelical decaying turbulence}
\begin{figure}
\includegraphics{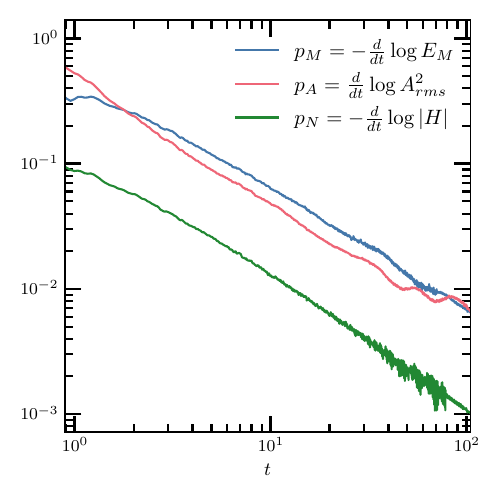}
\includegraphics{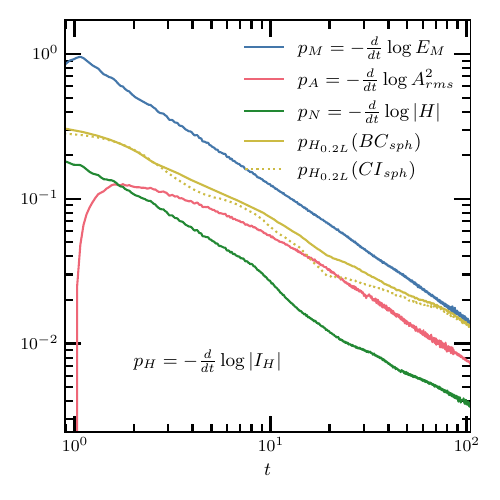}
\caption{Decay rates from 3D helical (upper panel) and nonhelical runs (lower panel). The integral $I_H$ was evaluated using both methods of box-counting (BC) and correlation-integral (CI) evaluated in a spherical (sph) region with radius $R=0.2L$. The decay rate of $\bra{\AAA^2}$ is lower than that of $I_H$ in the nonhelical case, indicating that former is the better conserved quantity.}
\label{decayrates}
\end{figure}
In BZL21, the study of 3D nonhelical decaying turbulence to investigate the possibility 
of magnetic reconnection driving the system was motivated by the similarity of the numerical 
result of $E_M \sim t^{-1}$ between 2D and 3D. In the 2D case, 
it is clear that the system is driven by reconnection. 
Additionally, vector-potential squared is not only ideally conserved
but also better conserved than magnetic energy in the limit of large Lundquist numbers. 
One can do a similar calculation as we did in section~\ref{helicalcase} using \Eq{decaylaw} 
and the constraint obtained from the 
conservation of vector-potential squared or anastrophy, 
\EQ
E_M k_p^{-2} \sim const.,
\label{anastrophycons}
\EN 
to obtain the scaling law, $E_M \sim t^{-1}$. 

In BZL21 it was claimed that conservation of anastrophy in \Eq{anastrophycons}, holds in 3D as well. To this end, BZL21 showed numerically that vector-potential squared is better conserved than magnetic energy in 3D. 
However, this result was overlooked in \cite{hosking_reconnection-controlled_2021} with the claim, "this will be true for any decay satisfying $ B^\alpha L \sim const.$ for any $\alpha > 0$", 
where $L$ is a proxy for $1/k_p$ or $L_{int}$. 
We explictly compare in a later subsection the rate of decay of various quantities and are able to make critical conclusions that go beyond the considerations specified above.
%Interestingly, vector potential squared does not decay at all in the case of helical decaying MHD turbulence, which corresponds to the case of $\alpha=2$.  
%In the helical case, $E_M \sim t^{-4/7}$ and thus $E_M k_p^{-2} \sim t^{4/7}$. Indeed, we find that, both in 2.5D and 3D helical runs, the vector-potential squared grows as a power law. 

In \citet{hosking_reconnection-controlled_2021}, they instead introduced a new quantity, termed the Saffman helicity invariant, 
\EQ
I_H = \int \bra{h(\xx)h(\xx+\rr)} d\rr,
\label{shi}
\EN
where $\bra{}$ represents an ensemble average
and claimed that this is the invariant which governs the 
the scaling laws in the nonhelical case. Their idea here is that 
a realistic system of MHD turbulence will contain helical fields of both polarities, resulting in zero net helicity. The integral $I_H$ 
which measures the correlation in the magnetic helicity density, is expected to be ideally conserved and drive the nonhelical decaying system at large Lundquist numbers. Using this, they 
obtain the following power law, $E_M \sim t^{-1.18}$, which is actually quite different from $t^{-1}$
measured in past numerical simulations \citep{zrake_inverse_2014,  berera_magnetic_2014, brandenburg_nonhelical_2015, reppin_nonhelical_2017, bhat_inverse_2021}. 

We have multiple reasons to conclude that $I_H$ is not the underlying integral constraining nonhelical turbulence decay. We delineate these below. 

\subsubsection{Decay of the ideal invariants/conserved quantities}

\begin{figure}
    \includegraphics{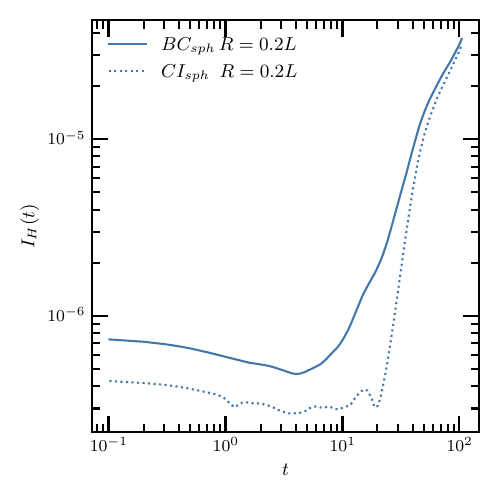}
    \caption{Evolution of the integral $I_H$ is evaluated by box-counting (BC) and correlation-integral (CI) methods in a spherical region with radius $R=0.2L$ in the 3D helical run with $1024^3$ resolution and $S=800$.}
    \label{IHgrowth}
\end{figure}

In the lower panel of \Fig{decayrates}, we show the rate of decay of magnetic energy, vector potential squared $\bra{\AAA^2}$, $I_H$ and the total 
magnetic helicity, $H_M$. We find that while $H_M$ is the slowest to change as expected, the next slowest is, in fact, $\bra{\AAA^2}$. Here, we have computed $I_H$ using the same algorithms as provided in \cite{zhou_scaling_2022},
available in \textsc{Pencil-Code}. In both methods of computing $I_H$ at $R=0.2L$, its rate of change is initially smaller than 
that of magnetic energy but larger compared to $\bra{\AAA^2}$. Eventually, the gap between the rate of change of $I_H$ and that of the magnetic energy reduces, and this is likely because $k_p^{-1} \gtrsim R$. 
Thus, we find that $I_H$ is not much slower than magnetic energy after around $t\sim 20$ and thus not as robustly 
conserved as $\bra{\AAA^2}$. We would like to further critically analyze the role of $I_H$ in decaying MHD turbulence, which we do in the next subsection. As expected, in the upper panel of \Fig{decayrates}, $H_M$ has the lowest decay rate. Notably, $\bra{\AAA^2}$ here actually grows instead of decaying (as we previously mentioned).

Another interesting thing to note is that in the helical case itself, $I_H$ does not decay, but grows with time as can be seen in \Fig{IHgrowth}. This reveals that the integral $I_H$ encodes the correlation scale of magnetic helicity. As the inverse transfer occurs, the integral scale shifts to larger values, indicating growth of helical structures to larger scales, which is also reflected in $I_H$. 

\subsubsection{Strength of helicity}

\begin{figure}
\begin{flushright}
\includegraphics{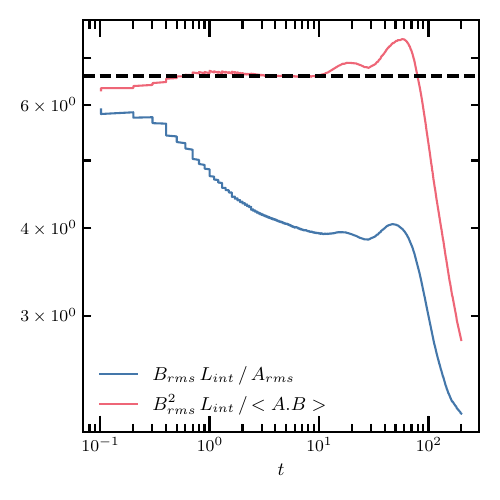}\\
\includegraphics{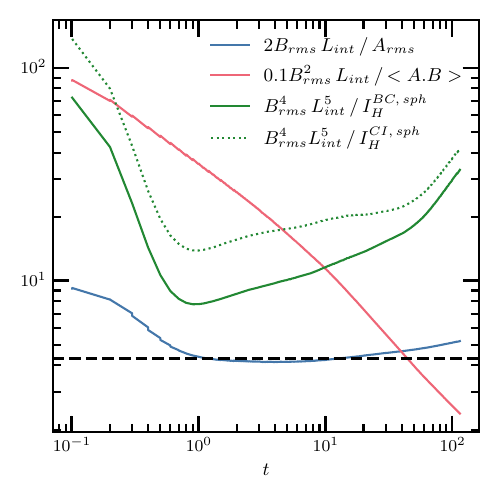}
\end{flushright}
\caption{Scalings from 3D ($1024^3$, $S=800$) runs. Upper panel: helical case, lower panel: nonhelical case. The ratio of $B_{rms}^4L_{int}^5/I_H$ (or equivalently, $E_M^2k_p^{-5}/I_H$) $\gg 1$ indicating low strength in magnetic helicity in individual structures of opposite polarities.}
\label{helstrength}
\end{figure}

The integral in \Eq{shi} can be calculated in two different ways \citep{zhou_scaling_2022}. 
The one that is used in study of its invariance properties in Fig. 5 in \cite{hosking_reconnection-controlled_2021} 
is to assume ensemble averages are equivalent to volume averages, where the averages are performed over volumes of size $R$, 
\EQA
I_H(R) &=& \frac{1}{V_R} \left( \int_{V_R} d^3r \int_{V_R} d^3x ~h(x) h(x+r) \right)  \nonumber \\ 
&\equiv& \frac{1}{V V_R} \int d^3x \left( \int_{V_R} d^3r ~h(x+r) \right)^2  
\ENA
The above is called the box counting method in \citet{zhou_scaling_2022}. 
The second way to calculate the integral is to retain the inner correlation integral, 
\EQ
I_H(R) = \frac{1}{V} \left( \int d^3r \int_{V_R} d^3x ~h(x) h(x+r) \right).
\EN
The way $I_H(R,t)$ varies with $ R $ or $ t $ in these two ways of calculation is similar.  This is because $R$ is restricted to 
the range, $L_{int} \ll R \ll L$, where $L$ is the system size. 

An important difference between magnetic helicity or vector-potential squared integrals and 
the $I_H$ integral is that in the former case, the magnetic structures are assumed to be mostly  
limited to a narrow range in wavenumber $k \sim k_p$ (here $1/k_p \sim L_{int}$). Hence, the integral 
over the volume is superficial, and thus, the scaling does not include a $L_{int}^3$ factor. 
However, in the latter case, the integral over the volume is necessary to 
beget the scaling of $E_M^2 k_p^{-5} \sim const.$. This complicates the scaling relation
as a simple dependence on a set of scales close to integral scale or the peak wavenumber $k_p$ is not well justified. 

However, our main criticism regarding the $I_H$ integral is related to the strength of the helicity in the smaller structures. The strength of helicity is related to whether the 
given magnetic field structure is fully helical. 
The realizability condition requires, $ k\vert H(k)\vert/2 \le M(k)$. 
When $k\vert H(k)\vert/2 = M(k)$, the magnetic fields on that scale or 
wavenumber are fully helical. In helical decaying MHD turbulence, 
the reason we can apply the constraint in \Eq{maghelcons} 
is because the fields in the system
are fully helical to begin with, which is the case in all the helical simulations 
in our paper as well as the previous ones. In fact, during the early part of the evolution, 
the fields remain fully helical, as we have seen already in \Fig{label-c}, 
which shows that even though the nonhelical part grows, it is smaller compared to $\Brms^H$ by more than an order of magnitude. However, the gap between the two curves keeps decreasing monotonically. 

Thus, in the fully helical case, we can indeed assume 
$\bra{\AAA \cdot \BB} \sim E_M L_{int} \sim const.$ is a good constraint. 
Similarly, in the nonhelical case, $I_H \sim E_M^2 L_{int}^5 \sim $ const. could be considered to robustly 
constrain the system evolution only if the individual magnetic structures of opposite polarities are fully helical. 
%Obtaining fully helical magnetic field topologies in astrophysics has been found to be quite uncommon \citep{}. 
Our simulations show in the lower panel of \Fig{helstrength} that indeed the strength of helicity is expectedly small in nonhelical turbulence and thus $E_M^2 k_p^{-5}/I_H \gg 1$ (or $B_{rms}^4 L_{int}^5/I_H \gg 1$). This ratio (here $I_H$ is measured at $R=0.2L$) 
remains larger than an order of magnitude (if we consider the dashed curve) compared to a factor of $\sim 2$ in the case 
of the ratio $B_{rms} L_{int}/A_{rms}$. 
 In the helical run as shown in the upper panel of \Fig{helstrength}, the ratio $B_{rms}^2 L_{int}/H_M \sim 2\pi$ (due to the factor in $L_{int}$), remains quite flat indicating the validity of usage of $H_M$ to constrain the system. Note that the curves corresponding to the ratios of $B_{rms}^2 L_{int}/H_M$ (in upper panel)
and $B_{rms} L_{int}/A_{rms}$ (lower panel) have initial amplitude around $6$ due to the $2\pi$ factor in $L_{int}$. In the 
case of $B_{rms}^4 L_{int}^5/I_H$, the volume cancels out and only $(2\pi)^2$ remains (due to squaring of helicity), which should have led to an initial amplitude of $~40$ if the structures were fully helical, but we get $100$ instead (as opposed to no such extra factors in the case of anastrophy here or magnetic helicity in the fully helical case). In fact the ratio involving magnetic helicity, $B_{rms}^2 L_{int}/H_M$, is as large as $\sim 1000$ initially 
given that the system is nonhelical.

In \Fig{helstrength}, both curves corresponding to $I_H$ and anastrophy dip initially in the transient phase. While the latter dips to a factor of $2$ and remains flat from around $t=1$ to $t=40$, 
the former touches a value of $10$ at $t=1$ and then increases again. Thus, we find that $B_{rms}^4 L_{int}^5$ is not robustly represented by $I_H$ due to the lack of strength in helicity. 

\begin{figure}
\begin{flushright}
\includegraphics{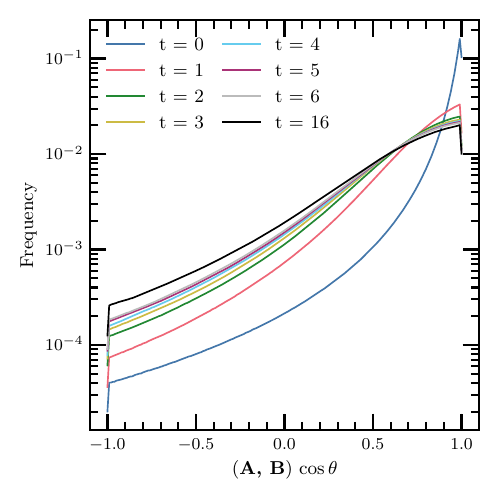}\\
\includegraphics{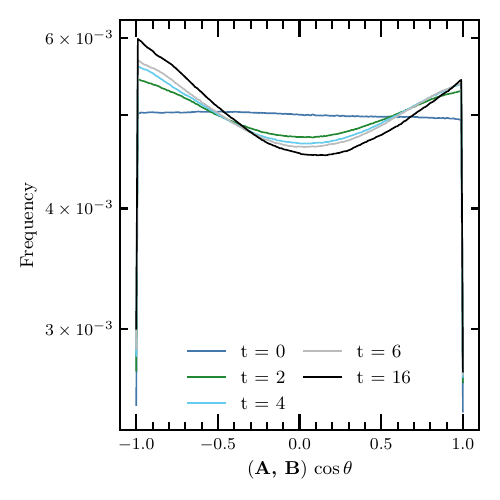}
\end{flushright}
\caption{Histograms of the cosine angle between the vector potential 
and the magnetic field from 3D ($1024^3$, $S=800$) runs. Upper panel: is the helical case showing a peak at $+1$ with frequency of $\sim 0.1$ and a sharp drop from there indicating that most of domain has fully helical fields (at all times). Lower panel: is the nonhelical case which is flat at $t=0$. Peaks at values of $\pm 1$ show up subsequently  but with an incremental factor of only $1.18$ indicating most of the domain is still largely fully nonhelical.}
\label{cosinehist}
\end{figure}
Our argument gains additional support from \Fig{cosinehist}. Here we present 
histograms of the cosine angle between the vector potential 
and the magnetic field, offering insights into their alignment. 
For fully helical fields, we obtain $\cos{\theta}=1$ and for 
fully nonhelical fields, $\cos{\theta}=0$.
In the helical case of decaying MHD turbulence, as shown in the upper panel of \Fig{cosinehist}, 
the magnetic fields are initialized to be random and fully helical. 
As expected, nearly 90\% of the total grid points (or counts) 
fall within the bins of $\cos{\theta} \in [0.8,1]$, and a steep drop from the peak is observed. 
Consequently, the portion of the domain containing nonhelical fields is 
minimal, and this trend persists as the system evolves.

On the other hand, the lower panel in \Fig{cosinehist} shows the histogram 
for the nonhelical case. In this case, the field is initially  
random but with no helicity. Hence, the $t=0$ curve is flat. 
According to the theory put forward by \cite{hosking_reconnection-controlled_2021}, 
the fully nonhelical fields should 
relax on a fast time scale, leaving the system with only helical fields. 
This decay should have manifested with peaks at both $\cos{\theta}=1$ and $\cos{\theta}=-1$ 
and a sharp drop. However, we find that there is a development of such peaks 
on a long time scale (compared to the Alfv\'en scale), and this growth in 
amplitude is only a factor of about 1.18, which is insignificant. The growth 
of such a peak mainly happens because it is easier for diffusion to kill 
fields which are less helical. Even at $t=16$, the percentage of points which 
lie within the set $\cos{\theta} \in [-1,-0.8] \cup [0.8, 1]$ is only about 25\%.
Additionally, we have checked that if we were to condition the histogram by the magnitude of the field i.e., we consider only those 
gridpoints where the field exceeds a threshold, say, $f\Brms$, where $f\ge 1$, the peaks are larger by only an incremental factor of about $\sim 2$ or so.
Thus, we show that in the nonhelical system, the strength of helicity is low. Therefore, a helicity-based integral such as $I_H$ likely cannot constrain the evolution 
of the system. 

Another crucial point to underscore is that even though magnetic helicity $H_M$ is well-conserved 
in also the nonhelical case (as evident in the lower panel of \Fig{decayrates}), we cannot use it to 
constrain the evolution of nonhelical system because the strength of helicity is insignificant. 
In a similar vein, $I_H$ may function as a well-conserved quantity (with a slightly lower decay rate 
compared to magnetic energy) \citep{zhou_scaling_2022}, but its effectiveness in constraining the evolution of a nonhelical system remains questionable.  

\subsubsection{Strictly 2D case of nonhelical decay}
\begin{figure}
    \centering
    \includegraphics{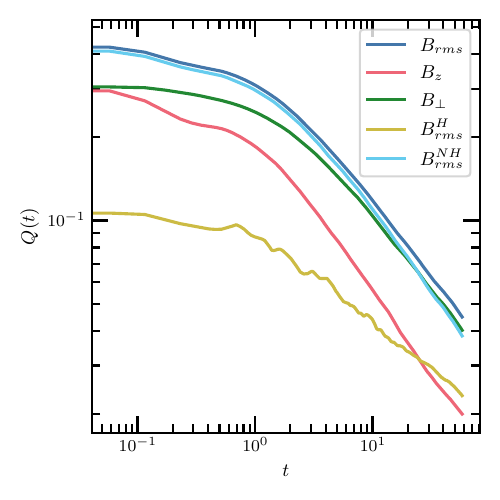}
    \caption{Evolution of the components of $\bm{B}$ for the nonhelical 2.5D case.}
    \label{fig:nonhel 2.5D bt}
\end{figure}
The third reason for concluding that $I_H$ is not important in decaying turbulence 
is related to the observation that, in both 2.5D 
and 3D, the helical and nonhelical cases show the same, decay in time, power laws. 
In the helical case, it is obviously the magnetic helicity conservation which plays the  critical role 
in constraining the system's evolution in both 2.5D and 3D. 
Similarly, in the nonhelical case, 
we expect the same conserved quantity (thus leading to the same results) to constrain evolution in both 2.5D 
as well as 3D. A further interesting observation is that, in the nonhelical case, the same decay power law of $E_M \sim t^{-1}$
holds in the strictly 2D case as well ($B_z=A_x=A_y=0$ throughout the simulation). In the 2D case (as opposed to 2.5D), magnetic helicity is 
not well-defined. Thus, any integral based on magnetic helicity is ruled out, including the 
Saffman helicity integral, $I_H$. Evolution plots from strictly 2D simulations with $E_M \sim t^{-1}$ 
were already seen in Fig. 1 of BZL21. There, we also confirmed that the strictly 2D simulations were  
similar to 2D simulations using the reduced-MHD model in \citet{zhou_magnetic_2019}. Thus, only vector-potential squared can 
explain the decay power law in the 2D case. And its influence extends to the 3D case as 
well, similar to magnetic helicity in the helical case. 

A further piece of evidence in favor of vector-potential squared $\bra{\AAA^2}$ is seen in \Fig{fig:nonhel 2.5D bt}. Here, we show different components of the magnetic field in a 2.5D nonhelical decaying turbulence system. We find that the system relaxes such that $B_z$ decays on a faster time-scale, making the in-plane component the dominant one. Such a behaviour could manifest also in 3D locally
around sites of reconnection (where most of the action takes place). Thus, we find it
is not only $\bra{\AAA_{2D}^2}$ (see SM of \citet{brandenburg_nonhelical_2015}) but even $\bra{\AAA^2}$ which is conserved.

\subsubsection{Local anisotropy}
\label{localaniso}
The question, then, is why there is a match between 2D (or 2.5D) and 3D results. 
Such a question was already posed in BZL21. The 2D-3D match has turned out to be more 
general, extending to the helical case as well. We continue to think that a quasi-two-dimensionalization
is involved, in the sense that the individual magnetic reconnection events seem to manifest in a 2D fashion 
within the 3D domain. In the reconnection literature, a quasi-two-dimensionalization manifests 
typically in the presence of a guide field, i.e. the reconnection rates are similar between a 2D case and a 3D case with a guide field. We resort to the well-known idea that local mean fields in the system 
can provide a local guide field for small-scale fields leading to local anisotropy \citep{cho_anisotropy_2000}. 
We study such possibilities using Minkowski functionals in the next section. 
~~\\

%-------------------------------------------------------------------------------------
\section{Quantifying magnetic field anisotropy using Minkowski Functionals}
Minkowski functionals (MFs) are tools which can be used to find the morphology of structures in a scalar field. It can help identify whether the isosurfaces in a given dataset are planar or filamentary, etc. The system of interest in this paper is magnetically dominated decaying MHD turbulence, which does not contain a guide field. Thus, it is not obvious if there can be anisotropy in the system, at least locally. But from our analysis in previous sections, we have provided ample evidence which suggests the possibility of quasi-two-dimensionalization of the 3D system at the relevant scales where magnetic reconnections occur. This likely includes the integral length scale (or the scales corresponding to the peak wavenumber $k_p$) and the smaller scales. Since we posit that reconnection in the 3D system manifests in a 2D fashion, from the previous literature, we gather that the reconnecting structures need to be influenced by a local guide field or mean field. Depending on the physics and the presence of local guide/mean fields, the magnetic structures can be rendered anisotropic in their morphology. We would like to use Minkowski functionals to detect and measure this anisotropy. 

We can test for quasi-two-dimensionalization by quantifying the spatial correlation of magnetic fields near reconnection sites in the computational domain. In such a quasi-2D scenario, the magnetic fields are expected to be correlated preferentially along one direction over the other two. That is, the fields should be anisotropic locally around the reconnection sites. 

\begin{figure}
    \centering
    \includegraphics[width=1\linewidth]{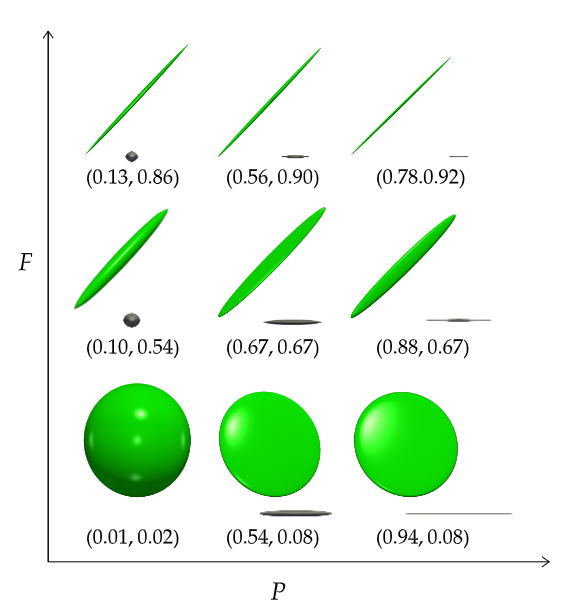}
    \caption{Known as the Blaschke diagram, we show the morphology of different spheroids as values of $P$ and $F$ varies along the abscissa and the ordinate respectively.}
    \label{blaschke}
\end{figure}
We first identify reconnection sites and then detect the magnetic fields around these sites. The necessary condition for reconnection to occur is the presence of high current density (see \cite{priest_mhd_2016} for sufficient conditions). Around a given high current density site, we analyze the correlation of magnetic fields by assessing the morphology of isonorm surfaces of $\vert \BB \vert$.  
The information regarding the shape of the isonorm surfaces is extracted using Minkowski Functionals (MFs). \citet{mecke_robust_1994} first introduced MFs in cosmology, and these were later employed by many others for morphological feature extraction (Refer to \cite{sheth} for a complete review of MFs). The execution of Minkowski functionals we employ is in Python from \citet{sheth}, who use the shape finders introduced by  \cite{sahni_shapefinders_1998}. The four Minkowski functionals are defined as,
\begin{equation}
    \label{minkowski functionals definition}
    \begin{split}
    V_0 = \iiint dV &\quad V_1 = \frac{1}{6}\iint dS\\ 
    V_2 = \frac{1}{6\pi} \iint\left(\kappa_1 + \kappa_2\right)dS &\quad V_3 = \frac{1}{4\pi}\iint \kappa_1\kappa_2 \,dS
\end{split}
\end{equation}
where $\kappa_1$ and $\kappa_2$ are the two principal curvatures. The volume and surface integrals, respectively, in the equations above, are specified over closed isosurfaces for a given field. We can extract three length scales from $V_0, V_1, V_2$ and $V_3$. The typical thickness, width, and length derived from (\ref{minkowski functionals definition}) are, 
\begin{equation}
    \label{thickness, width, length}
    T = \frac{V_0}{2V_1}, \quad W = \frac{2V_1}{\pi V_2}, \quad L = \frac{3V_2}{4V_3}.
\end{equation}
These three length scales are ordered such that \(l_1 = \text{max}(T,W,L), l_2 = \text{mid}(T,W,L), l_3 = \text{min}(T,W,L)\). Next, we can obtain planarity and filamentarity for a given isosurface structure as follows,
\begin{equation}
\label{planarity-filamentarity}
    P = \frac{l_2-l_3}{l_2+l_3}, \quad F = \frac{l_1 - l_2}{l_1+l_2}.
\end{equation}
To both validate our implementation of the MFs and provide an understanding of the morphology of structures in a phase space governed by $P$ and $F$, we show the Blaschke diagram in \Fig{blaschke}. Here, the abscissa varies over planarity $P$, and the ordinate varies over filamentarity $F$. It is clear from \Eq{planarity-filamentarity} that the range of $P$ and $F$ is $[0,1]$, where zero means no planarity or no filamentarity. 
In \Fig{blaschke}, we show how the shapes change depending on the values of ($P, F$). 
We have a sphere in the lower left corner of the plot that changes to a plate as $P$ increases (with $F$ constant) in the lower row but changes to a rugby ball-like shape as $F$ increases (with $P$ constant) in the left column. 

Further, we have also performed the MF analysis separately for fully developed hydrodynamic turbulence and confirmed the expected results. This can be found in appendix~\ref{hydroMF}. 

Since magnetic reconnection occurs near points where there is a significant value of current density, we implement an algorithm to obtain isolated clusters of points with the current density value $\vert \JJ \vert$ exceeding a threshold of $0.7 \vert \JJ\vert_{\text{max}}$. In each such cluster, we designate the point with the highest value as the representative point. We designate the current density isosurface with isovalue of $0.5  \vert\JJ\vert_{\text{max, cluster}}$ as our fiducial current sheet for that cluster. 

\begin{figure}
    \centering
    \includegraphics[width=0.48\textwidth]{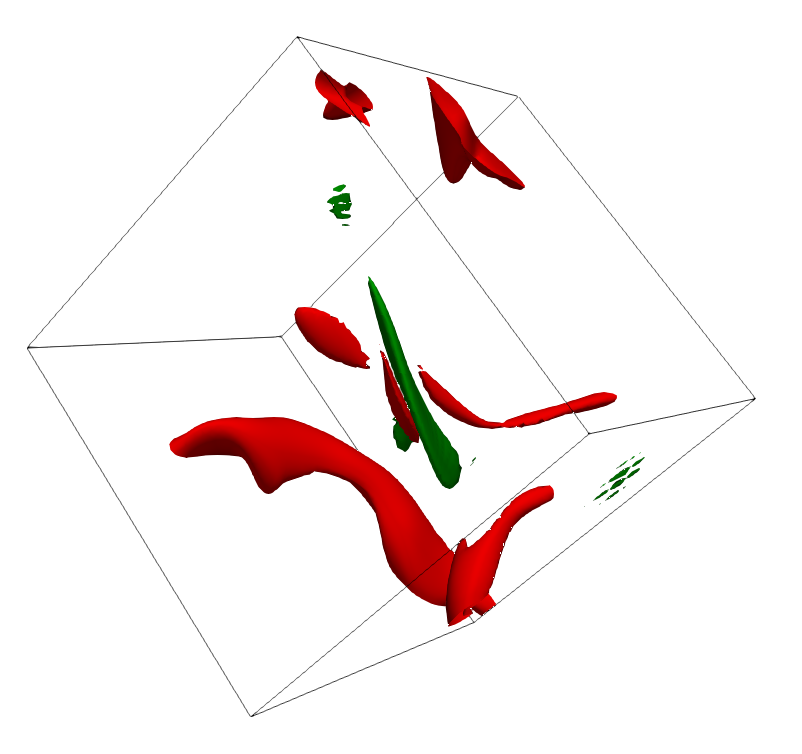}
    \caption{Zoomed in version of isosurfaces of the current density $\vert \JJ \vert$ and the magnetic field $\vert \BB \vert$ are shown in green and red respectively from the nonhelical run of $1024^3$ with $S=800$. The domain size shown here is $72^3$.}
    \label{fig:nonhel jb iso}
\end{figure}
\begin{figure}
    \centering
    \includegraphics[width=0.48\textwidth]{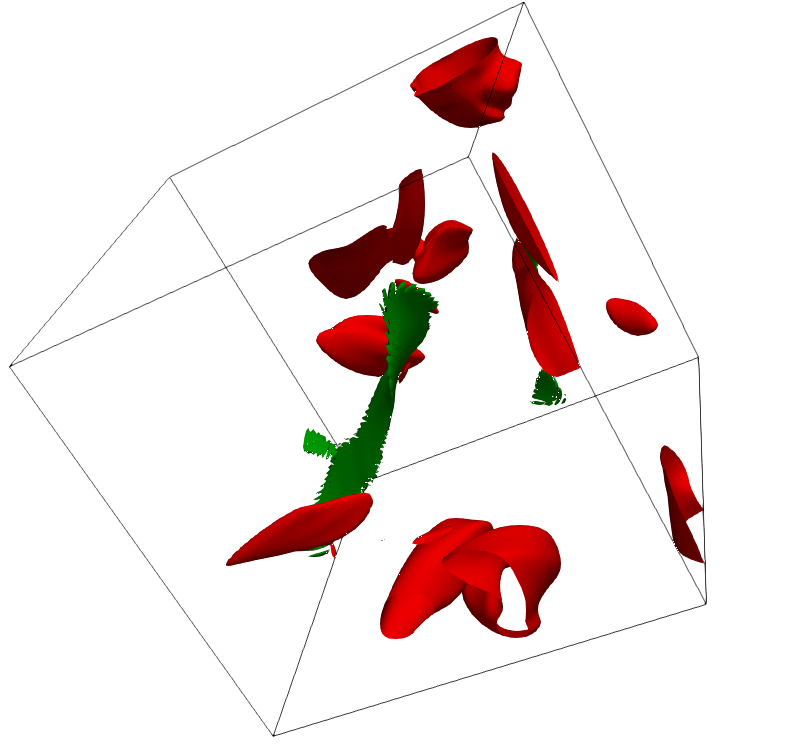}
    \caption{Zoomed in version of isosurfaces of the current density $\vert \JJ \vert$ and the magnetic field $\vert \BB \vert$ are shown in green and red, respectively, from the fully helical run of $1024^3$ with $S=800$. The domain size shown here is $116^3$.}
    \label{fig:hel jb iso}
\end{figure}

\begin{figure*}
    \centering
     \includegraphics[width=0.45\textwidth]{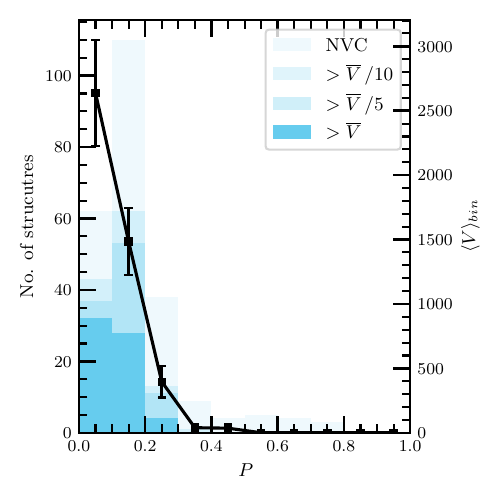}
     \includegraphics[width=0.45\textwidth]{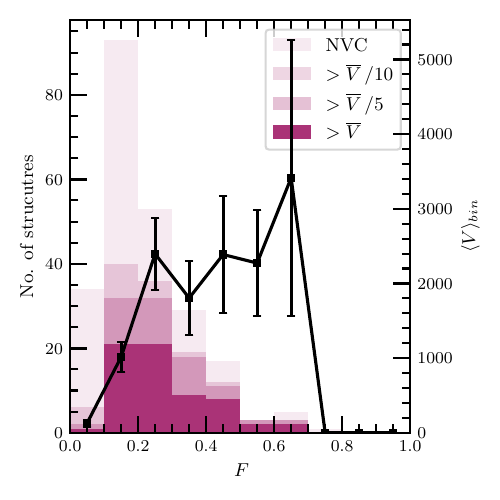}
    \caption{\textbf{Nonhelical:} Histogram for planarity (left panel) and filamentarity (right panel) of all the \textit{reconnecting magnetic field structures} in the nonhelical $1024^3$ simulation with $S=800$, at $t=5$. Also shown in the black squares is the average volume in each bin. NVC stands for no-volume-cutoff. The darker shades represent statistics for structures considered above a certain volume threshold denoted by $\overline{V}/n$ and $n=1$, $5$ or $10$. The tendency of large structures to have smaller planarity and larger filamentarity is clear from the graph. The same conclusion can also be reached by progressively considering structures with larger and larger volumes, as we have shown in this plot.}
    \label{histograms-nonhel}
\end{figure*}
\begin{figure*}%[h!]
     \centering
     \includegraphics{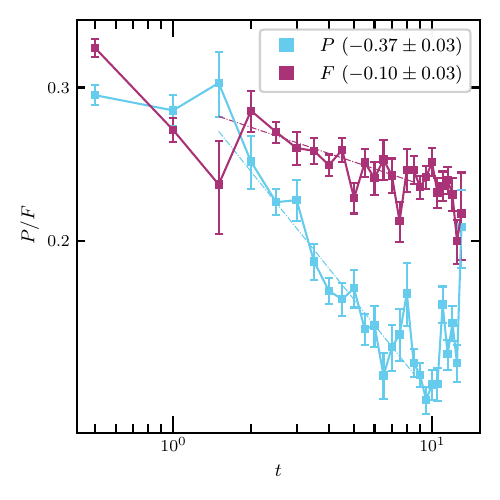}
     \includegraphics{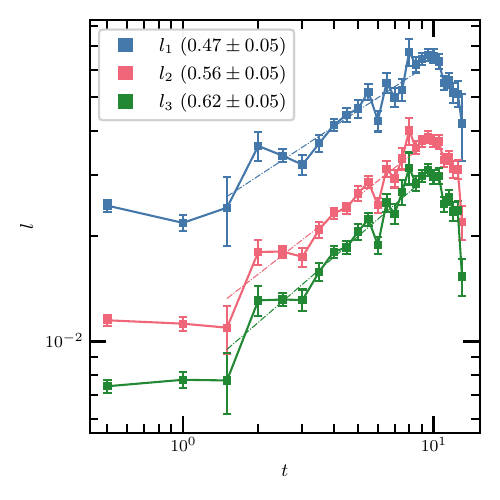}
     \caption{\textbf{Nonhelical:} Time evolution of $P$ and $F$ in the early stages of the decay (when the $B_{rms}$ curve behaviour is close to the theoretical prediction of $t^{-0.5}$), is shown in the left panel.
     The time evolution of length scales of \textit{reconnecting magnetic field} isosurfaces is shown in the right panel. The different dashed curves are the least square fit, and they indicate a scaling close to the expected $l \sim t^{1/2}$.}
     \label{fig:nonhel l and pf evolve}
 \end{figure*}
 
Next, we would like to obtain a set of magnetic field isosurfaces close to this current sheet. 
So, we employ a range of isovalues to do this. To determine this range of isovalues, consider the following 
scheme.
Around a \textit{current sheet} (in an isolated cluster), $\vert \BB \vert$ is calculated along rays emanating from the representative point. Once we obtain the $\vert \BB \vert$ profile around the representative point,  we find the maximum and minimum value of $\vert \BB \vert$ nearest to the point. Then we use these extrema values to determine a range of isovalues of $\vert \BB \vert$ given by,  
\begin{multline}\vert \BB \vert_{\text{max}} - \frac{i}{5} \times \left(\frac{\vert \BB \vert_{\text{max}}-\vert \BB \vert_{\text{min}}}{4}\right) \\ \forall i \in {1,2,3,4,5}.\end{multline}
Thereafter, Minkowski functionals for $\vert \JJ \vert$ and $\vert \BB \vert$ surfaces are calculated, and their evolution with time is reported. The details  of algorithm for MFs implementation and the scheme we described briefly for obtaining isosurfaces is available in 
Appendix~\ref{algoMF} and \ref{detailed MF steps}.
                                 
\subsection{Results for Minkowski functionals}

 For an illustration, we show in \Figs{fig:nonhel jb iso}{fig:hel jb iso} the isosurfaces of an isolated current density structure corresponding to a possible reconnection site and also isosurfaces of the magnetic fields nearby that are potentially participating in the reconnection. We calculate the MFs for such structures of current density and the magnetic field throughout the domain of the $1024^3$ runs of $S=800$ in both helical and nonhelical cases. We extract the quantities $l_1$, $l_2$, $l_3$ and the corresponding $P$ and $F$ values. Below, we study their distributions (histograms) and evolution in time.

\subsubsection{Nonhelical case}
In \Fig{histograms-nonhel}, we show the histograms of the measured planarity $P$ and filamentarity $F$ at a given time, $t=5$ from the $1024^3$ run of $S=800$. 
In these charts, NVC (no-volume-cutoff) denotes the histogram for the complete set of structures. The darker shades correspond to histograms derived from a subset where only structures with a volume surpassing a specific threshold are considered. The darkest shade represents the subset where only structures with a volume exceeding the average value $\overline{V}$ are considered.
The planarity histogram (left panel of \Fig{histograms-nonhel}) peaks within the bin of $0.1$--$0.2$. With subsets of larger volume structures, the histograms progressively shrink towards lower values of $P$. This indicates the tendency of more voluminous structures to have smaller planarity. In the case of filamentarity histogram  (right panel of \Fig{histograms-nonhel}), the peak is within the bin of $0.2$--$0.3$. Here, as we limit the set to having more voluminous structures, the average value of $F$ shifts to the higher side.  
On the whole, the structures are more filamentary than planar. These values indicate that the structures feature between the sphere and the middle ellipsoid in the first column of the Blaschke diagram in \Fig{blaschke}. Thus, their geometry is akin to prolate spheroids, which are more filamentary than they are planar (unlike the oblate spheroid). This kind of anisotropy inherent in the magnetic structure is likely reflective of the local anisotropy within the system.

Next, we study the evolution of the morphology of the structures. In the left panel of \Fig{fig:nonhel l and pf evolve}, we show the evolution of the average value of $P$ and $F$ calculated for the magnetic fields. Firstly, both $P$ and $F$ tend to decrease. This can be understood as the increasing effect of diffusion, which tends to even out the structures. While the decrease in $F$ is very small ($\sim t^{-0.1}$ given by a straight line fit as mentioned in the legend), the decrease in $P$ is larger. Here, the $P$ and $F$ values were calculated by taking the average of the values obtained from all the individual isosurfaces(detailed steps are in Appendix \ref{detailed MF steps}). These average $P$ and $F$ values do not have to necessarily match with that resulting from the average values of $l_1$, $l_2$ and $l_3$ (let us call them $P^*$ and $F^*$), and indeed they don’t. However, the trends match, though the indices of power law decay are different, i.e. $P^*\sim t^{-0.25\pm0.03}$ and $F^*\sim t^{-0.2\pm0.04}$.  

\begin{figure*}
    \centering
    \includegraphics[width=0.45\textwidth]{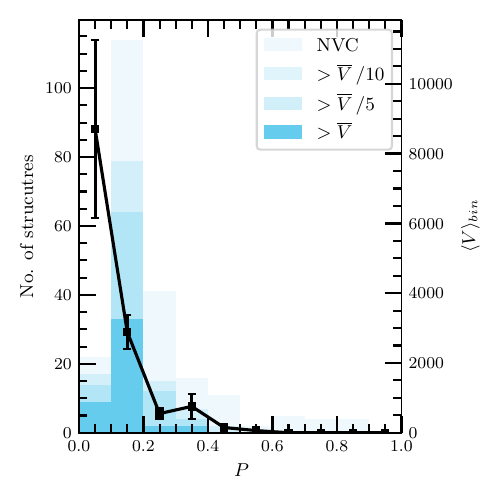}
     \includegraphics[width=0.45\textwidth]{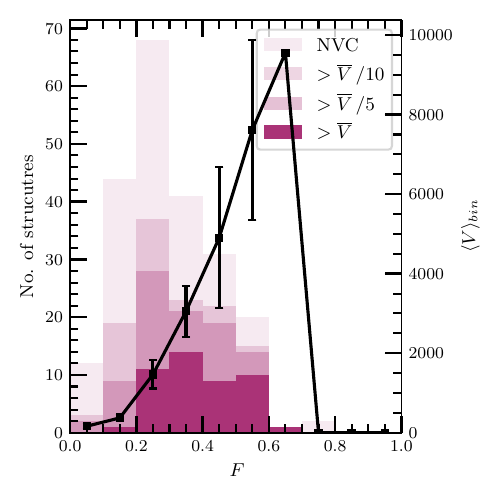}
    \caption{\textbf{Helical:} Histogram for planarity (left panel) and filamentarity (right panel) of all the \textit{reconnecting magnetic field structures} in the helical $1024^3$ simulation with $S=800$, at $t=5$. Also shown in the black squares is the average volume in each bin. NVC stands for no-volume-cutoff. The darker shades represent statistics for structures considered above a certain volume threshold denoted by $\overline{V}/n$ and $n=1$, $5$ or $10$. The tendency of large structures to have smaller planarity and larger filamentarity is clear from the graph. The same conclusion can also be reached by progressively considering structures with larger and larger volumes, as we have shown in this plot.}
    \label{histograms-hel}
\end{figure*}
 \begin{figure*}%[h!]
 	\centering
 	\includegraphics{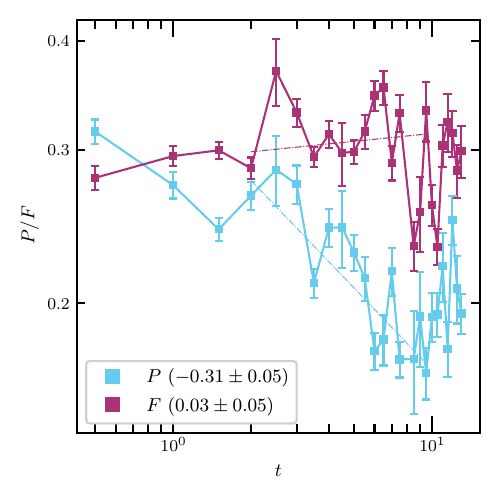}
 	\includegraphics{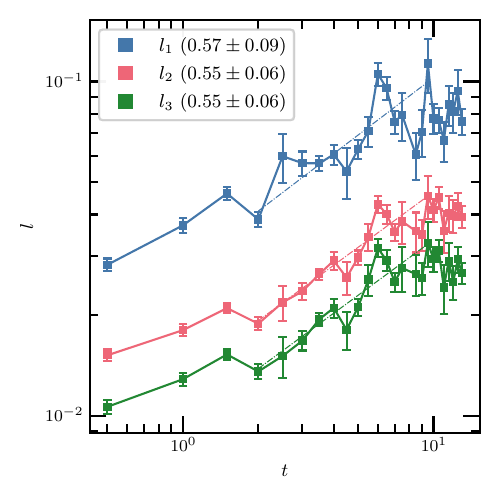}
 	\caption{\textbf{Helical:} Time evolution of $P$ and $F$ in the early stages of the decay (when the $B_{rms}$ curve behaviour is close to the theoretical prediction of $t^{-2/7}$), is shown in the left panel.
     The time evolution of length scales of \textit{reconnecting magnetic field} isosurfaces is shown in the right panel. The different dashed curves are the least square fit, and they indicate a scaling close to the expected $l \sim t^{4/7}$.}
 	\label{fig:hel bb l evolve}
 \end{figure*}
The behaviour of $P^*$/$P$ and $F^*$/$F$ is roughly consistent with the power law growth of  $l_1$, $l_2$ and $l_3$ as shown in the right panel of \Fig{fig:nonhel l and pf evolve}. The thin dashed lines indicate a straight line fit to the evolution curves, and we infer the respective power-law indices. 
It can be seen that indeed $l_2$ and $l_3$ increase faster than $l_1$. This can be understood as the effect of diffusion, which acts more efficiently on smaller scales. Thus, the indices increase progressively as we go from $l_1$ to $l_3$. Otherwise, the values of the power law indices (within the error bars) are close to the theoretically expected value of 0.5 (more so for $l_1$ and $l_2$ than $l_3$, which is most affected by diffusion). 
It's important to note that the accuracy of the straight line fit is influenced by the chosen time range. Nevertheless, we demonstrate that a fit closely aligning with the theoretical expectation is not unreasonable, considering the estimated error bars.

\subsubsection{Helical case}

Next, we consider the iso-surfaces in the decaying helical MHD turbulence. 
In \Fig{histograms-hel}, we once again show the histograms of the measured planarity $P$ and filamentarity $F$ at a given time, $t=5$ from the helical $1024^3$ run of $S=800$.
The peak of the planarity histogram, similar to that of the nonhelical case, peaks within the bin of $0.1-0.2$. However, the filamentarity histogram peaks at a larger value within the bin of $0.3-0.4$ for more voluminous structures. Here again, as we restrict to a subset of structures with larger volumes, the averages shift to lower values for $P$ and higher values for $F$. As in the nonhelical case, the structures are more filamentary (even more so) than planar. 

In the left panel of \Fig{fig:hel bb l evolve}, we show the evolution of the average value of $P$ and $F$ to assess how the morphology of the structures changes. Here, $F$ is nearly constant as time evolves, whereas the value of $P$ shows a decrease (similar to the nonhelical case) with time. In the right panel of \Fig{fig:hel bb l evolve}, we show the evolution of $l_1$, $l_2$ and $l_3$ and we find here the decay of $l_2$ and $l_3$ are not all that much faster than $l_1$ (unlike the nonhelical case). We can chalk up this behaviour and constancy of $F$ to the fact that helical structures tend to be more resilient to diffusion \citep{bhat_resilience_2014}. Here again, we find that the power law indices from the straight line fit, as shown are close to the theoretically expected value of $\sim 0.57$.  

~~\\
\par
\begin{figure}
    \centering
    \includegraphics{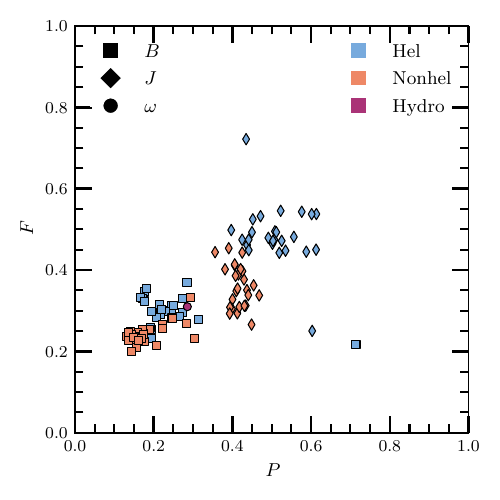}
    \caption{Blaschke plot of the combined helical, nonhelical, and hydrodynamic simulations. The $(P, F)$ of the vorticity of the forced hydrodynamic simulation has been given as a reference point against which the $(P, F)$ of the MHD simulations could be compared.
    Note that there is an outlier in the $(P, F)$ of the magnetic field in the helical simulation at $(0.71,0.22)$ which occurs at $t=8$. This is because there was only one current sheet above the threshold, so we could not obtain enough statistics. We ignore this data point in the evolution plots. }
    \label{fig:hel nonhel hydro pf compare}
\end{figure}

 MF analysis was also performed for the current density structures. In \Fig{fig:hel nonhel hydro pf compare}, we show the phase space Blaschke plot for $P$ and $F$ values from all the different cases of helical MHD, nonhelical MHD and hydrodynamics (no magnetic fields). Here, the values of ($P, F$) for the current density structures (shown in a diamond shape) are around $\sim (0.53,0.5)$ in the helical case and $\sim (0.45, 0.37)$ in the nonhelical case. Thus, unlike the case of magnetic field isosurfaces, the current density structures appear between the bottom two structures in the middle column of the Blaschke diagram in \Fig{blaschke}. Thus, they are largely sheet-like, reaffirming the understanding of this system as a quasi-2D system with 2D-like reconnections.  

\section{A quasi-2D hierarchical merger model}

A hierarchical merger model was first proposed in \cite{zhou_magnetic_2019} to understand 
a 2D system is driven by island mergers due to coalescence instability. 
It was useful to derive scaling laws associated with the decay of magnetic energy 
and resulting inverse transfer. While BZL21 suggested that the 
3D decaying MHD turbulence is akin to the 2D system (also with similar scaling laws), 
but they didn't provide a 3D model to describe the system. A 3D model was proposed 
in \cite{zhou_multi-scale_2020}, but it was for a system with a uniform mean field 
or guide field. We find that their hierarchical model is incompatible with 
what we propose below. 

In our quasi-2D hierarchical merger model (Q2DHM model), interactions between magnetic structures happen 
pairwise. An interaction consists of magnetic reconnection at the interface, leading 
to a merger. Such a merger, then, conserves mass and magnetic flux 
or magnetic helicity (depending on the system). 
The resulting structures are then a size larger but with a lower strength of magnetic field. 
Thus, the two ingredients which were required to solve the Kolmogorov-type calculation 
using decay equation in \Eq{decaylaw} is available 
in the Q2DHM model as well : (i) the reconnection timescale being the driving timescale
for the system, (ii) a conserved quantity to constrain the relationship with the magnetic field 
and the integral scale in the problem. 
However, the advantage of the Q2DHM model over the decay equation in \Eq{decaylaw} is 
that it allows us to make the distinction 
between a 2D and a 3D system and explicitly see the manifestation of quasi-two-dimensionality. 

In the model, the mergers happen in discrete stages (given by $n$).
A key assumption is that at any given stage, there is 
uniformity in magnetic field structures, i.e. all of them have the same 
dimensions of $l_{1n}$, $l_{2n}$ and $l_{3n}$. 
The magnetic field in a structure is given by $B_n$. 

First, the merger of two structures conserves mass. Assuming the condition 
of incompressibility, we can write, 
\begin{equation}
2 l_{1_n} l_{2_n} l_{3_n} = l_{1_{n+1}} l_{2_{n+1}} l_{3_{n+1}} 
\label{masscons}
\end{equation}
where $n+1$ denotes the next stage (or the next generation). 
Thus, we have $l_{1_{n+1}}=2^{1/3}l_{1_n}$, $l_{2_{n+1}}=2^{1/3}l_{2_n}$ 
and $l_{3_{n+1}}=2^{1/3}l_{3_n}$. 

Second, we demand flux conservation. However, for a system with 
quasi-two-dimensionality, the structures reconnect  
on a plane guided by a local mean field, then the conservation of flux 
can be given by, 
\begin{equation}
B_n l_{i_n} \sim const.   
\label{fluxcons}
\end{equation}
where the $i$ subscript can assume $1,2$ or $3$ depending on the 
plane of reconnection. Thus, we have $B_{n+1}=2^{-1/3}B_n=2^{-n/3}B_0$, 
where $B_0$ is the initial field. 
We can now calculate the Lundquist number at the $n+1$ stage, 
\begin{equation}
S_{n+1}= \frac{B_{n+1} l_{i_{n+1}}}{\eta} = \frac{2^{-1/3}B_n 2^{1/3}l_{i_n}}{\eta} = S_n
\label{Sevol}
\end{equation}
Since each merger is driven by reconnection, the corresponding timescale is given 
by the inverse of the reconnection rate. Since we assume the 2D steady-state model of 
Sweet-Parker for reconnection, we obtain, 
\begin{equation}
\tau_{{rec}_{n+1}} = S_{n+1}^{1/2} \tau_{A_{n+1}} = 2^{2/3} \tau_{A_n} = 2^{2n/3} \tau_0
\label{taurec}
\end{equation}
The time taken to reach the $n$th generation can be 
approximated to be $t_n\approx 2^{(2/3)n}\tau_0$, for large $n$. 
Thus we have time $\tilde{t}=t_n/\tau_0=2^{(2/3)n}$. 
Now, we can eliminate $n$ to obtain time dependence of various quantities, 
\begin{eqnarray}
&B=B_0 \tilde{t}^{-1/2}, \quad l_1=l_{1_0} \tilde{t}^{1/2} \\ \nonumber
&l_2=l_{2_0} \tilde{t}^{1/2}, \quad l_3=l_{3_0} \tilde{t}^{1/2}
\label{nonhelscalings}
\end{eqnarray}
Thus, we recover the scalings which otherwise were previously derived using equation \Eq{decaylaw}. 
Note that if we do not assume the quasi-two-dimensionality for \Eq{fluxcons} and 
instead used $B_n l_{i_n}l_{j_n} \sim const.$, we would have obtained very different scalings. 

In the helical case, instead of \Eq{fluxcons}, the more important topological constraint 
is derived from magnetic helicity. Thus, for each structure, we consider 
\begin{equation}
B^2_n l_{i_n} \sim const.   
\label{helcons}
\end{equation}
Here, then we have $B_{n+1}=2^{-1/6}B_n=2^{-n/6}B_0$. 
This will affect the the scaling of $S$, 
\begin{equation}
S_{n+1}= \frac{B_{n+1} l_{i_{n+1}}}{\eta} = \frac{2^{-1/6}B_n 2^{1/3}l_{i_n}}{\eta} = 2^{1/6} S_n
\label{Sevol2}
\end{equation}
Consequently, instead of \Eq{taurec}, we obtain, 
\begin{equation}
\tau_{{rec}_{n+1}} = S_{n+1}^{1/2} \tau_{A_{n+1}} = 2^{7/12} \tau_{A_n} = 2^{7n/12} \tau_0
\label{taurec2}
\end{equation}
Note that for calculating $\tau_{rec}$ in both cases, we use the 2D Sweet-Parker model. 
We can then, as before, eliminate $n$ to obtain time dependence of various quantities, 
\begin{eqnarray}
&B=B_0 \tilde{t}^{-2/7}, \quad l_1=l_{1_0} \tilde{t}^{4/7} \\ \nonumber
&l_2=l_{2_0} \tilde{t}^{4/7}, \quad l_3=l_{3_0} \tilde{t}^{4/7}
\label{helscalings}
\end{eqnarray}
Thus, we recover the scalings observed in our simulations, which are also consistent with the Kolmogorov-type calculation.

\section{Solutions based on self-similarity}

Another approach to recovering the scaling laws for decaying turbulence involves leveraging the self-similarity property of the MHD equations, as initially proposed by \citet{olesen}. However, the self-similarity scaling suggested by \citet{olesen} resulted in all the physics being determined by the initial MHD spectrum $k^\alpha$. This limitation was identified by \citet{brandenburg_classes_2017}, who proposed an improved form for the self-similarity scaling. Such a spectrum is given by, 
\EQ
E(k, t) = \xi(t)^{-\beta} \phi(k \xi(t))
\label{selfsim}
\EN
where $\xi(\kappa)$ is a universal function, $\xi$ is similar to $L_{int}$ and $\beta$ governs the envelope 
traced by the evolving peak wavenumber. Writing down the power law exponents as, 
\EQ
p=-\frac{d \ln E_M}{d \ln t},~ q= \frac{d \ln L_{int}}{d \ln t},  
\label{exp}
\EN
we obtain, 
\EQ
p= (\beta + 1)q.
\label{2ndrel}
\EN
From the above \Eq{2ndrel}, we can infer the $\beta$ given the conserved quantity. For example, in the case of magnetic helicity, we have $E_M L_{int} \sim const.$, which leads to $p=-q$ and, thus, the corresponding $\beta=0$. However, we need another constraint to recover the exponents $p$ and $q$ themselves. 

Traditionally, the scaling solutions assume the inherent timescale in the problem is given by $t_{\rm decay}=L_{int}/\sqrt{E_M}$ or the Alfv\'enic timescale. For such a relationship, we have $1=q+p/2$.
However, in the magnetically dominated decaying turbulence, the reconnection timescale becomes important. And then we obtain instead $t_{\rm decay}=S^{1/2}L_{int}/\sqrt{E_M}$ which leads to,
\EQ
1= \frac{1}{4}(6q+p).
\label{3rdrel}
\EN
Now, with \Eqs{2ndrel}{3rdrel}, we can recover the power law exponents. 
For the case of fully helical turbulence with $\beta=0$, we obtain $p=4/7$ and $q=4/7$ as before in \Eq{helscalings}. In the nonhelical case, $\beta=1$ for anastrophy conservation. With $\beta=1$, we recover $p=1$ and $q=1/2$ as before in \Eq{nonhelscalings}.
%----------------------------------------------------------------------------------------------------------------------------------------
\section{Discussions}

\paragraph*{Dynamical timescale is the reconnection timescale.}
This is best seen in the strict 2D case, where one can explicitly track magnetic islands merging via reconnection. In 3D, it is hard to track reconnecting structures. However BZL21 showed not only curve-collapse in 3D runs with reconnection timescale taken into account, they found a match in 2D and 3D scalings, which we continue to find in helical case as well in this paper. Further, BZL21 study of shell-to-shell energy transfer functions was also consistent with a reconnection-driven system.
These results, along with that in \citet{hosking_reconnection-controlled_2021}, seem to rule out Alfv\'en timescale as the relaxation/decay timescale. However, many other studies continue to regard Alfv\'en timescale as the decay timescale and thus obtain scaling exponents for magnetic energy that are different but very close to that obtained for reconnection timescale such as $-10/9$ and $-2/3$ for nonhelical and helical cases respectively (as compared to $-1$ and $-4/7$) \citep{zhou_scaling_2022,brandenburg_hosking_2023,brandenburg_inverse_2023}. Another concern is whether many of these studies, which run high-resolution simulations at all, resolve reconnection physics (i.e. current sheet widths).

\paragraph*{Necessity of high Lundquist number and its connection to quasi-two-dimensionality}
The simulation results of inverse transfer in nonhelical MHD 
decaying turbulence posed a puzzle upon their emergence, given the absence 
of magnetic helicity in this system \citep{zrake_inverse_2014, brandenburg_nonhelical_2015}. Notably, this is a high magnetic Reynolds number 
or high $S$
(available only at a higher resolution) result, which explains its non-detection 
in many of the earlier simulations of nonhelical MHD decaying turbulence 
\citep{christensson_inverse_2001}. Thus, a comprehensive understanding of this system should duly consider this important factor.  
At low $S$ or in a low-resolution simulation, the timescales of operation for microscopic 
diffusion can get sufficiently short such that any effect of local anisotropy can be 
rendered ineffective. However, a more significant issue arises when the occurrence of magnetic reconnection is impeded at sites of large current density due to non-negligible diffusion (aided by turbulence).
One may inquire about the implications for systems that are not nonhelical. Similar considerations should be taken into account in that scenario as well. Indeed, we observe a steeper decay in lower resolution simulations, which does not necessarily align with the theoretical calculation of $E_M \sim t^{-4/7}$. Thus, whether nonhelical or helical, in both cases, the system at large resolution and high Reynolds numbers tends to be more quasi-two-dimensional. Consequently, it is not surprising when the relevant reconnection time-scale corresponds to a 2D model, i.e., Sweet-Parker reconnection.

\paragraph*{A uniquely MHD result}
This quasi-two-dimensionalization leads to exact same decay laws in both 2D and 3D, 
which is a uniquely MHD result. Indeed, in hydrodynamics, the 2D and 3D decay laws
are different, with different conserved quantities at play. Such a 2D-3D similarity 
is not to be expected in hydrodynamics unless there is an agent for bringing about 
quasi-two-dimensionalization, which in the MHD case is naturally manifested by 
large magnetic field strength (even if the fields are of turbulent nature and reside at smaller scales). 
In hydrodynamic turbulence, such an agent could be strong rotation, and then the relevant 
conserved quantities will change back to that relevant in 2D as expected \citep{yakhot_largescale_1987}. 

\paragraph*{Local mean fields}
We attribute the quasi-two-dimensionalization to the presence of 
non-zero local mean magnetic fields. 
 In the standard theories for steady-state magnetically dominated 
 MHD turbulence (either weak or strong), a mean-field is required as 
 the explicit source for anisotropy. The dynamics, then, are described 
 along the mean field separately from across it. Depending on whether the 
 turbulence is weak or strong, connections are made between 
 the dynamics perpendicular and parallel to the mean field \citep{tobias_mhd_2011}. 
 In our system, there is no apriori global mean field (or uniform field), 
 however, we postulate that as the strong random fields reconnect, 
 a net local mean field 
 can arise to influence the dynamics locally. 
 This concept aligns with the conventional understanding in standard MHD turbulence theories that, ultimately, the local mean field is important (despite the presence of a uniform mean field) \citep{cho_anisotropy_2000}. 
However, in this study, we have refrained from directly calculating the local mean fields. The reason is that it is unclear what the scale size for local 
mean field's region of influence needs to be for it to render the reconnection quasi-2D. 

\paragraph*{3D reconnection}
Studies on 3D reconnection are sparse. Most studies either discuss various 
kinds of manifestations of reconnection sites or arising current instabilities 
(for e.g. kink instability \citep{pontin_three-dimensional_2011, landi_three-dimensional_2008, oishi_self-generated_2015}). While it is well known that a guide field 
or mean-field renders the ensuing reconnection in the system effectively 2D. Another manifestation of guide field reconnection is when two flux tubes interact at an angle 
\citep{linton_reconnection_2001}. However,  
a complete and systematic characterization of such systems is required for understanding 
within a larger context of turbulent systems. 

\paragraph*{Prandtl number dependence} 
Magnetic reconnection has been shown to be less efficient at higher magnetic Prandtl number $P_M=\nu/\eta$ \citep{Comisso_Grasso_Waelbroeck_2015}. The associated reconnection timescale is shown to increase with $P_M$, $\tau_{rec} \propto (1+P_M)^{1/2}$. While this has been used to propose that the early universe magnetic fields retain significant amplitude to explain the fields observed in cosmic voids \citep{hosking_cosmic-void_2023}, it is not clear that such a relation should continue to hold at large values of $P_M$. Some studies suggest that inverse transfer itself gets inefficient with increasing $P_M$
 \citep{armua_parameter_2023, reppin_nonhelical_2017}.

\paragraph*{Anisotropy measurement}
Previous studies that have measured anisotropy have used structure function 
calculation, taking into account the local mean field in the calculation \citep{cho_anisotropy_2000}.
However, given the uncertainty in the required coherence scale for the mean field, 
we are unable to use similar techniques. Instead, we use Minkowski functionals 
to calculate the characteristic length scales of the magnetic field structures. 
All the three length scales $l_1, l_2$ and $l_3$ we obtain from this analysis 
are very different, leading to non-zero significant planarity and filamentarity. 
While the difference between $l_1$ and $l_2$ or $l_3$ can be attributed to local 
mean-field, that between $l_2$ and $l_3$ could be due cross helicity considerations 
and dynamic alignment \citep{boldyrev_dynamic_2009}. We have not studied these in this work and leave it 
for a future investigation.

\paragraph*{Issue of gauge}
Already, magnetic helicity is a tricky topological quantity to deal with 
depending on the nature of the system under study, and there are many issues to consider 
in its calculation. Depending on the boundary conditions and other ingredients 
like stratification, large-scale shear, etc, there can be magnetic helicity fluxes 
which can change the conservation properties dramatically. Then there is the issue 
related to gauge, which relates to the meaning and interpretation of the results 
in a system with nontrivial (which are less academic and more realistic) boundaries. 
But as long as one is in a periodic system, even though local helicity fluxes maybe present due to some inhomogeneities, results can be considered to be robust for the total magnetic helicity.

Similar considerations can be extended to the quantity of 
vector-potential squared. This quantity is physically connected to the concept 
of flux-freezing. In 2D, magnetic flux conservation is directly related to 
the invariance of anastrophy or vector-potential squared. 
In our decaying nonhelical MHD turbulence case, 
anastrophy is particularly useful given that the system displays 
quasi-two-dimensionalization. In the absence of mean magnetic helicity, 
this quantity displays slower decay time scales compared to magnetic energy 
and this is fairly nontrivial as, in the case of non-zero magnetic helicity, 
anastrophy tends to increase with time. In other contexts 
related to condensates in Yang-Mills theories, the minimization 
of this quantity w.r.t the topology of the magnetic field, is 
associated with a gauge condition which is invariant. 
This is used as a probe of phase transition \citep{gubarev_significance_2001}. 
Thus, we find that understanding the 3D system with anastrophy is not unjustified 
on the grounds of gauge arguments as then those would equally apply to 
a more commonly used 3D quantity like magnetic helicity density. 

\paragraph*{3D Hierarchical merger models}
In our Quasi-2D Hierarchical Merger (Q2DHM) model, we posit that due to 
quasi-two-dimensionalization, the magnetic flux conservation constraint 
is for the in-plane magnetic field (w.r.t reconnection). This is consistent 
with the requirement of vector-potential squared or anastrophy conservation 
we discussed above. As mentioned previously, our model is not compatible 
with that presented in \cite{zhou_multi-scale_2020} as our first constraint 
of mass conservation leads to \Eq{masscons}, which is not the case in theirs. Further, 
they draw constraints from strong MHD turbulence of critical balance. 
The nature of turbulence in this system is not yet fully understood due to 
the lack of probes on the nature of the local mean fields. 
However, as mentioned in BZL21, the slope of $k^{-2}$ in the magnetic spectrum 
is more compatible with that of weak turbulence. 

\section{Conclusions}

In a previous paper, we had shown that magnetic reconnection is important 
in decaying nonhelical MHD turbulence. In this work, we showed that the same 
is true of decaying fully helical MHD turbulence as well. 
Total magnetic helicity, a conserved quantity in the limit of 
high Lundquist numbers, is used to constrain the evolution of the helical case. 
We found that the predicted theoretical power law decay exponent pans out in 
both 2.5D and 3D simulations. 
We critically analyzed two possibilities of ideal invariants 
that could constrain evolution in the nonhelical case, (i) anastrophy or vector potential 
squared (which is ideally conserved in 2D or 2.5D), $\bra{\AAA^2}$ (ii) the integral based on helicity fluctuations, $I_H$ (which is undefined in strict 2D). 
We showed that $\bra{\AAA^2}$ has a lower decay rate as compared to $I_H$ or the magnetic energy $\bra{\BB^2}/2$. Next, we argued that the assumption of the nonhelical 
system being a composite of fully helical structures of opposite polarities (where the fully nonhelical part decays away on fast timescales) is not valid. We provided evidence for this with (i) a histogram of the cosine angle between vector potential and magnetic field in \Fig{cosinehist}, where the concentration at and around $\pm 1$ is low, and (ii) we showed that the ration of $E_M^2 k_p^{-5}/I_H \gg 1$. Our final contention was that the same nonhelical behaviour is also observed in strict 2D case, where $I_H$ is basically undefined. 

Next, we investigated the possibility of quasi-two-dimensionalization, which can explain the similarity seen in power law decay exponents between 2D (or 2.5D) and 3D cases. To this end, we performed a Minkowski functional analysis for the \textit{reconnecting} magnetic fields. We found that both helical and nonhelical fields have a planarity $P \sim 0.1$--$0.2$ and filamentarity $F\sim 0.2$--$0.3$, indicating the existence of local anisotropy. 
Also, we showed that the fit to power law evolution curves of the characteristic scales, $l_1$, $l_2$ and $l_3$ displays exponents that are close to the theoretically expected values of $0.5$ and $0.57$ in nonhelical and helical cases respectively. 

We provided a Quasi-2D Hierarchical Merger (Q2DHM) model based on mass conservation and either magnetic flux conservation (in the nonhelical case) or helicity conservation (in the helical case). In this model, we recover the observed scalings only when we consider (i) a 2D manifestation of flux-freezing consistent with vector-potential squared conservation in the plane of reconnection, (ii) and Sweet-Parker model (a 2D model) for reconnection.

\section*{Acknowledgements}

SD, CA and PB are thankful to Prof. Kandaswamy Subramanian for his insightful comments and discussions. We are thankful also to the reviewers for their challenging reports which resulted in further strengthening of our results. 
SD expresses gratitude to ICTS for hosting her in the Long Term Visiting Students Program.  
CA thanks Vinay Kumar and Rajarshi Chattopadhyay for their valuable help and discussions.
We acknowledge project RTI4001 of the Dept. of Atomic Energy, Govt. of India.
SD acknowledges support from the US Dept. of Energy (grant number DE-SC0018266).
The simulations were performed on the ICTS HPC cluster \textit{Contra}. 

%----------------------------------------------------------------------------------------------------------------------------------------

%\section{Bibliography}  
%%%%%%%%%%%%%%%%%%%% REFERENCES %%%%%%%%%%%%%%%%%%

% The best way to enter references is to use BibTeX:

\bibliographystyle{mnras}
\bibliography{anisoturb} % if your bibtex file is called example.bib

\begin{thebibliography}{}
\makeatletter
\relax
\def\mn@urlcharsother{\let\do\@makeother \do\$\do\&\do\#\do\^\do\_\do\%\do\~}
\def\mn@doi{\begingroup\mn@urlcharsother \@ifnextchar [ {\mn@doi@} {\mn@doi@[]}}
\def\mn@doi@[#1]#2{\def\@tempa{#1}\ifx\@tempa\@empty \href {http://dx.doi.org/#2} {doi:#2}\else \href {http://dx.doi.org/#2} {#1}\fi \endgroup}
\def\mn@eprint#1#2{\mn@eprint@#1:#2::\@nil}
\def\mn@eprint@arXiv#1{\href {http://arxiv.org/abs/#1} {{\tt arXiv:#1}}}
\def\mn@eprint@dblp#1{\href {http://dblp.uni-trier.de/rec/bibtex/#1.xml} {dblp:#1}}
\def\mn@eprint@#1:#2:#3:#4\@nil{\def\@tempa {#1}\def\@tempb {#2}\def\@tempc {#3}\ifx \@tempc \@empty \let \@tempc \@tempb \let \@tempb \@tempa \fi \ifx \@tempb \@empty \def\@tempb {arXiv}\fi \@ifundefined {mn@eprint@\@tempb}{\@tempb:\@tempc}{\expandafter \expandafter \csname mn@eprint@\@tempb\endcsname \expandafter{\@tempc}}}

\bibitem[\protect\citeauthoryear{Armua, Berera  \& Calderón-Figueroa}{Armua et~al.}{2023}]{armua_parameter_2023}
Armua A.,  Berera A.,   Calderón-Figueroa J.,  2023, \mn@doi [Physical Review E] {10.1103/PhysRevE.107.055206}, 107, 055206

\bibitem[\protect\citeauthoryear{Batchelor \& Taylor}{Batchelor \& Taylor}{1949}]{batchelor_role_1949}
Batchelor G.~K.,  Taylor G.~I.,  1949, \mn@doi [Proceedings of the Royal Society of London. Series A. Mathematical and Physical Sciences] {10.1098/rspa.1949.0007}, 195, 513

\bibitem[\protect\citeauthoryear{Berera \& Linkmann}{Berera \& Linkmann}{2014}]{berera_magnetic_2014}
Berera A.,  Linkmann M.,  2014, \mn@doi [Physical Review E] {10.1103/PhysRevE.90.041003}, 90, 041003

\bibitem[\protect\citeauthoryear{Bhat \& Subramanian}{Bhat \& Subramanian}{2013}]{bhat_fluctuation_2013}
Bhat P.,  Subramanian K.,  2013, \mn@doi [Monthly Notices of the Royal Astronomical Society] {10.1093/mnras/sts516}, 429, 2469

\bibitem[\protect\citeauthoryear{Bhat, Blackman  \& Subramanian}{Bhat et~al.}{2014}]{bhat_resilience_2014}
Bhat P.,  Blackman E.~G.,   Subramanian K.,  2014, \mn@doi [Monthly Notices of the Royal Astronomical Society] {10.1093/mnras/stt2402}, 438, 2954

\bibitem[\protect\citeauthoryear{Bhat, Ebrahimi, Blackman  \& Subramanian}{Bhat et~al.}{2017}]{bhat_evolution_2017}
Bhat P.,  Ebrahimi F.,  Blackman E.~G.,   Subramanian K.,  2017, \mn@doi [Monthly Notices of the Royal Astronomical Society] {10.1093/mnras/stx1989}, 472, 2569

\bibitem[\protect\citeauthoryear{Bhat, Zhou  \& Loureiro}{Bhat et~al.}{2021}]{bhat_inverse_2021}
Bhat P.,  Zhou M.,   Loureiro N.~F.,  2021, \mn@doi [Monthly Notices of the Royal Astronomical Society] {10.1093/mnras/staa3849}, 501, 3074

\bibitem[\protect\citeauthoryear{Blandford, Yuan, Hoshino  \& Sironi}{Blandford et~al.}{2017}]{blandford_magnetoluminescence_2017}
Blandford R.,  Yuan Y.,  Hoshino M.,   Sironi L.,  2017, \mn@doi [Space Science Reviews] {10.1007/s11214-017-0376-2}, 207, 291

\bibitem[\protect\citeauthoryear{Boldyrev, Mason  \& Cattaneo}{Boldyrev et~al.}{2009}]{boldyrev_dynamic_2009}
Boldyrev S.,  Mason J.,   Cattaneo F.,  2009, \mn@doi [The Astrophysical Journal] {10.1088/0004-637X/699/1/L39}, 699, L39

\bibitem[\protect\citeauthoryear{Brandenburg}{Brandenburg}{2023}]{brandenburg_hosking_2023}
Brandenburg A.,  2023, \mn@doi [Journal of Plasma Physics] {10.1017/S0022377823000028}, 89, 175890101

\bibitem[\protect\citeauthoryear{Brandenburg \& Kahniashvili}{Brandenburg \& Kahniashvili}{2017}]{brandenburg_classes_2017}
Brandenburg A.,  Kahniashvili T.,  2017, \mn@doi [Physical Review Letters] {10.1103/PhysRevLett.118.055102}, 118, 055102

\bibitem[\protect\citeauthoryear{Brandenburg, Kahniashvili  \& Tevzadze}{Brandenburg et~al.}{2015}]{brandenburg_nonhelical_2015}
Brandenburg A.,  Kahniashvili T.,   Tevzadze A.~G.,  2015, \mn@doi [Physical Review Letters] {10.1103/PhysRevLett.114.075001}, 114, 075001

\bibitem[\protect\citeauthoryear{Brandenburg, Sharma  \& Vachaspati}{Brandenburg et~al.}{2023}]{brandenburg_inverse_2023}
Brandenburg A.,  Sharma R.,   Vachaspati T.,  2023, \mn@doi [Journal of Plasma Physics] {10.1017/S0022377823001253}, 89, 905890606

\bibitem[\protect\citeauthoryear{Cho \& Vishniac}{Cho \& Vishniac}{2000}]{cho_anisotropy_2000}
Cho J.,  Vishniac E.~T.,  2000, \mn@doi [The Astrophysical Journal] {10.1086/309213}, 539, 273

\bibitem[\protect\citeauthoryear{Christensson, Hindmarsh  \& Brandenburg}{Christensson et~al.}{2001}]{christensson_inverse_2001}
Christensson M.,  Hindmarsh M.,   Brandenburg A.,  2001, \mn@doi [Physical Review E] {10.1103/PhysRevE.64.056405}, 64, 056405

\bibitem[\protect\citeauthoryear{Comisso, Grasso  \& Waelbroeck}{Comisso et~al.}{2015}]{Comisso_Grasso_Waelbroeck_2015}
Comisso L.,  Grasso D.,   Waelbroeck F.~L.,  2015, \mn@doi [Physics of Plasmas] {10.1063/1.4918331}, 22, 042109

\bibitem[\protect\citeauthoryear{Daughton, Roytershteyn, Karimabadi, Yin, Albright, Bergen  \& Bowers}{Daughton et~al.}{2011}]{daughton_role_2011}
Daughton W.,  Roytershteyn V.,  Karimabadi H.,  Yin L.,  Albright B.~J.,  Bergen B.,   Bowers K.~J.,  2011, \mn@doi [Nature Physics] {10.1038/nphys1965}, 7, 539

\bibitem[\protect\citeauthoryear{Gubarev, Stodolsky  \& Zakharov}{Gubarev et~al.}{2001}]{gubarev_significance_2001}
Gubarev F.~V.,  Stodolsky L.,   Zakharov V.~I.,  2001, \mn@doi [Physical Review Letters] {10.1103/PhysRevLett.86.2220}, 86, 2220

\bibitem[\protect\citeauthoryear{Hesse \& Cassak}{Hesse \& Cassak}{2020}]{hesse_magnetic_2020}
Hesse M.,  Cassak P.~A.,  2020, \mn@doi [Journal of Geophysical Research: Space Physics] {10.1029/2018JA025935}, 125, e2018JA025935

\bibitem[\protect\citeauthoryear{Hosking \& Schekochihin}{Hosking \& Schekochihin}{2021}]{hosking_reconnection-controlled_2021}
Hosking D.~N.,  Schekochihin A.~A.,  2021, \mn@doi [Physical Review X] {10.1103/PhysRevX.11.041005}, 11, 041005

\bibitem[\protect\citeauthoryear{Hosking \& Schekochihin}{Hosking \& Schekochihin}{2023}]{hosking_cosmic-void_2023}
Hosking D.~N.,  Schekochihin A.~A.,  2023, \mn@doi [Nature Communications] {10.1038/s41467-023-43258-3}, 14, 7523

\bibitem[\protect\citeauthoryear{{Kolmogorov}}{{Kolmogorov}}{1941}]{kolmogorov_dissipation_1997}
{Kolmogorov} A.~N.,  1941, Akademiia Nauk SSSR Doklady, \href {https://ui.adsabs.harvard.edu/abs/1941DoSSR..32...16K} {32, 16}

\bibitem[\protect\citeauthoryear{{Landau} \& {Lifshitz}}{{Landau} \& {Lifshitz}}{1975}]{LL1975}
{Landau} L.~D.,  {Lifshitz} E.~M.,  1975, {The classical theory of fields}

\bibitem[\protect\citeauthoryear{Landi, Londrillo, Velli  \& Bettarini}{Landi et~al.}{2008}]{landi_three-dimensional_2008}
Landi S.,  Londrillo P.,  Velli M.,   Bettarini L.,  2008, \mn@doi [Physics of Plasmas] {10.1063/1.2825006}, 15, 012302

\bibitem[\protect\citeauthoryear{Lesieur \& Ossia}{Lesieur \& Ossia}{2000}]{lesieur_3d_2000}
Lesieur M.,  Ossia S.,  2000, \mn@doi [Journal of Turbulence] {10.1088/1468-5248/1/1/007}, 1, 007

\bibitem[\protect\citeauthoryear{Lewiner, Lopes, Vieira  \& Tavares}{Lewiner et~al.}{2003}]{lewiner}
Lewiner T.,  Lopes H.,  Vieira A.~W.,   Tavares G.,  2003, \mn@doi [Journal of Graphics Tools] {10.1080/10867651.2003.10487582}, 8, 1–15

\bibitem[\protect\citeauthoryear{Linton, Dahlburg  \& Antiochos}{Linton et~al.}{2001}]{linton_reconnection_2001}
Linton M.~G.,  Dahlburg R.~B.,   Antiochos S.~K.,  2001, \mn@doi [The Astrophysical Journal] {10.1086/320974}, 553, 905

\bibitem[\protect\citeauthoryear{Lorensen \& Cline}{Lorensen \& Cline}{1987}]{lorensen}
Lorensen W.~E.,  Cline H.~E.,  1987, \mn@doi [ACM SIGGRAPH Computer Graphics] {10.1145/37402.37422}, 21, 163–169

\bibitem[\protect\citeauthoryear{Matthaeus \& Montgomery}{Matthaeus \& Montgomery}{1980}]{matthaeus_selective_1980}
Matthaeus W.~H.,  Montgomery D.,  1980, \mn@doi [Annals of the New York Academy of Sciences] {10.1111/j.1749-6632.1980.tb29687.x}, 357, 203

\bibitem[\protect\citeauthoryear{McKinney \& Uzdensky}{McKinney \& Uzdensky}{2012}]{mckinney_reconnection_2012}
McKinney J.~C.,  Uzdensky D.~A.,  2012, \mn@doi [Monthly Notices of the Royal Astronomical Society] {10.1111/j.1365-2966.2011.19721.x}, 419, 573

\bibitem[\protect\citeauthoryear{McPherron}{McPherron}{1979}]{mcpherron_magnetospheric_1979}
McPherron R.~L.,  1979, \mn@doi [Reviews of Geophysics] {10.1029/RG017i004p00657}, 17, 657

\bibitem[\protect\citeauthoryear{Mecke, Buchert  \& Wagner}{Mecke et~al.}{1994}]{mecke_robust_1994}
Mecke K.~R.,  Buchert T.,   Wagner H.,  1994, \mn@doi [Astronomy and Astrophysics] {10.48550/arXiv.astro-ph/9312028}, 288, 697

\bibitem[\protect\citeauthoryear{Mininni \& Pouquet}{Mininni \& Pouquet}{2013}]{mininni_inverse_2013}
Mininni P.~D.,  Pouquet A.,  2013, \mn@doi [Physical Review E] {10.1103/PhysRevE.87.033002}, 87, 033002

\bibitem[\protect\citeauthoryear{Moffatt, Kida  \& Ohkitani}{Moffatt et~al.}{1994}]{hydro_ref}
Moffatt H.~K.,  Kida S.,   Ohkitani K.,  1994, \mn@doi [Journal of Fluid Mechanics] {10.1017/S002211209400011X}, 259, 241–264

\bibitem[\protect\citeauthoryear{Narayan, Igumenshchev  \& Abramowicz}{Narayan et~al.}{2003}]{narayan_magnetically_2003}
Narayan R.,  Igumenshchev I.~V.,   Abramowicz M.~A.,  2003, \mn@doi [Publications of the Astronomical Society of Japan] {10.1093/pasj/55.6.L69}, 55, L69

\bibitem[\protect\citeauthoryear{Neronov \& Vovk}{Neronov \& Vovk}{2010}]{neronov_evidence_2010}
Neronov A.,  Vovk I.,  2010, \mn@doi [Science] {10.1126/science.1184192}, 328, 73

\bibitem[\protect\citeauthoryear{Oishi, Low, Collins  \& Tamura}{Oishi et~al.}{2015}]{oishi_self-generated_2015}
Oishi J.~S.,  Low M.-M.~M.,  Collins D.~C.,   Tamura M.,  2015, \mn@doi [The Astrophysical Journal Letters] {10.1088/2041-8205/806/1/L12}, 806, L12

\bibitem[\protect\citeauthoryear{Olesen}{Olesen}{1997}]{olesen}
Olesen P.,  1997, \mn@doi [Physics Letters B] {10.1016/S0370-2693(97)00235-9}, 398, 321

\bibitem[\protect\citeauthoryear{Parker}{Parker}{1983}]{parker_magnetic_1983}
Parker E.~N.,  1983, \mn@doi [The Astrophysical Journal] {10.1086/160636}, 264, 635

\bibitem[\protect\citeauthoryear{{Peebles}}{{Peebles}}{1980}]{Peebles1980}
{Peebles} P.~J.~E.,  1980, {The large-scale structure of the universe}

\bibitem[\protect\citeauthoryear{Pevtsov, Fisher, Acton, Longcope, Johns-Krull, Kankelborg  \& Metcalf}{Pevtsov et~al.}{2003}]{pevtsov_relationship_2003}
Pevtsov A.~A.,  Fisher G.~H.,  Acton L.~W.,  Longcope D.~W.,  Johns-Krull C.~M.,  Kankelborg C.~C.,   Metcalf T.~R.,  2003, \mn@doi [The Astrophysical Journal] {10.1086/378944}, 598, 1387

\bibitem[\protect\citeauthoryear{Pontin}{Pontin}{2011}]{pontin_three-dimensional_2011}
Pontin D.~I.,  2011, \mn@doi [Advances in Space Research] {10.1016/j.asr.2010.12.022}, 47, 1508

\bibitem[\protect\citeauthoryear{Pontin, Wilmot-Smith, Hornig  \& Galsgaard}{Pontin et~al.}{2011}]{pontin_dynamics_2011}
Pontin D.~I.,  Wilmot-Smith A.~L.,  Hornig G.,   Galsgaard K.,  2011, \mn@doi [Astronomy \& Astrophysics] {10.1051/0004-6361/201014544}, 525, A57

\bibitem[\protect\citeauthoryear{Porth, Komissarov  \& Keppens}{Porth et~al.}{2014}]{porth_three-dimensional_2014}
Porth O.,  Komissarov S.~S.,   Keppens R.,  2014, \mn@doi [Monthly Notices of the Royal Astronomical Society] {10.1093/mnras/stt2176}, 438, 278

\bibitem[\protect\citeauthoryear{Pouquet, Frisch  \& Léorat}{Pouquet et~al.}{1976}]{pouquet_frisch_léorat_1976}
Pouquet A.,  Frisch U.,   Léorat J.,  1976, \mn@doi [Journal of Fluid Mechanics] {10.1017/S0022112076002140}, 77, 321–354

\bibitem[\protect\citeauthoryear{Priest}{Priest}{2016}]{priest_mhd_2016}
Priest E.,  2016, in Gonzalez W.,  Parker E.,  eds, Astrophysics and {Space} {Science} {Library}, Magnetic {Reconnection}: {Concepts} and {Applications}.
Springer International Publishing, Cham, pp 101--142, \mn@doi{10.1007/978-3-319-26432-5_3}, \url {https://doi.org/10.1007/978-3-319-26432-5_3}

\bibitem[\protect\citeauthoryear{Rappazzo, Velli, Einaudi  \& Dahlburg}{Rappazzo et~al.}{2008}]{rappazzo_nonlinear_2008}
Rappazzo A.~F.,  Velli M.,  Einaudi G.,   Dahlburg R.~B.,  2008, \mn@doi [The Astrophysical Journal] {10.1086/528786}, 677, 1348

\bibitem[\protect\citeauthoryear{Reppin \& Banerjee}{Reppin \& Banerjee}{2017}]{reppin_nonhelical_2017}
Reppin J.,  Banerjee R.,  2017, \mn@doi [Physical Review E] {10.1103/PhysRevE.96.053105}, 96, 053105

\bibitem[\protect\citeauthoryear{Ruan, Xia  \& Keppens}{Ruan et~al.}{2020}]{ruan_fully_2020}
Ruan W.,  Xia C.,   Keppens R.,  2020, \mn@doi [The Astrophysical Journal] {10.3847/1538-4357/ab93db}, 896, 97

\bibitem[\protect\citeauthoryear{Sahni, Sathyaprakash  \& Shandarin}{Sahni et~al.}{1998}]{sahni_shapefinders_1998}
Sahni V.,  Sathyaprakash B.~S.,   Shandarin S.~F.,  1998, \mn@doi [The Astrophysical Journal] {10.1086/311214}, 495, L5

\bibitem[\protect\citeauthoryear{Seta, Bushby, Shukurov  \& Wood}{Seta et~al.}{2020}]{seta_saturation_2020}
Seta A.,  Bushby P.~J.,  Shukurov A.,   Wood T.~S.,  2020, \mn@doi [Physical Review Fluids] {10.1103/PhysRevFluids.5.043702}, 5, 043702

\bibitem[\protect\citeauthoryear{Sheth, Sahni, Shandarin  \& Sathyaprakash}{Sheth et~al.}{2003}]{sheth}
Sheth J.~V.,  Sahni V.,  Shandarin S.~F.,   Sathyaprakash B.~S.,  2003, \mn@doi [Monthly Notices of the Royal Astronomical Society] {10.1046/j.1365-8711.2003.06642.x}, 343, 22–46

\bibitem[\protect\citeauthoryear{Shukurov \& Subramanian}{Shukurov \& Subramanian}{2021}]{kandubook}
Shukurov A.,  Subramanian K.,  2021, Astrophysical {Magnetic} {Fields}: {From} {Galaxies} to the {Early} {Universe}.
Cambridge {Astrophysics}, Cambridge University Press, Cambridge, \mn@doi{10.1017/9781139046657}, \url {https://www.cambridge.org/core/books/astrophysical-magnetic-fields/231CE7B2753A87CEB4BFA55AE04412D1}

\bibitem[\protect\citeauthoryear{Subramanian}{Subramanian}{2016}]{subramanian_origin_2016}
Subramanian K.,  2016, \mn@doi [Reports on Progress in Physics] {10.1088/0034-4885/79/7/076901}, 79, 076901

\bibitem[\protect\citeauthoryear{Subramanian, Shukurov  \& Haugen}{Subramanian et~al.}{2006}]{subramanian_evolving_2006}
Subramanian K.,  Shukurov A.,   Haugen N. E.~L.,  2006, \mn@doi [Monthly Notices of the Royal Astronomical Society] {10.1111/j.1365-2966.2006.09918.x}, 366, 1437

\bibitem[\protect\citeauthoryear{Taylor}{Taylor}{1986}]{taylor_relaxation_1986}
Taylor J.~B.,  1986, \mn@doi [Reviews of Modern Physics] {10.1103/RevModPhys.58.741}, 58, 741

\bibitem[\protect\citeauthoryear{Ting, Matthaeus  \& Montgomery}{Ting et~al.}{1986}]{ting_turbulent_1986}
Ting A.~C.,  Matthaeus W.~H.,   Montgomery D.,  1986, \mn@doi [The Physics of Fluids] {10.1063/1.865843}, 29, 3261

\bibitem[\protect\citeauthoryear{Tobias, Cattaneo  \& Boldyrev}{Tobias et~al.}{2011}]{tobias_mhd_2011}
Tobias S.~M.,  Cattaneo F.,   Boldyrev S.,  2011, {MHD} {Dynamos} and {Turbulence}, \mn@doi{10.48550/arXiv.1103.3138}, \url {http://arxiv.org/abs/1103.3138}

\bibitem[\protect\citeauthoryear{Wiegand, Buchert  \& Ostermann}{Wiegand et~al.}{2014}]{wiegand_direct_2014}
Wiegand A.,  Buchert T.,   Ostermann M.,  2014, \mn@doi [Monthly Notices of the Royal Astronomical Society] {10.1093/mnras/stu1118}, 443, 241

\bibitem[\protect\citeauthoryear{Wilkin, Barenghi  \& Shukurov}{Wilkin et~al.}{2007}]{wilkin_magnetic_2007}
Wilkin S.~L.,  Barenghi C.~F.,   Shukurov A.,  2007, \mn@doi [Physical Review Letters] {10.1103/PhysRevLett.99.134501}, 99, 134501

\bibitem[\protect\citeauthoryear{Yakhot \& Pelz}{Yakhot \& Pelz}{1987}]{yakhot_largescale_1987}
Yakhot V.,  Pelz R.,  1987, \mn@doi [The Physics of Fluids] {10.1063/1.866294}, 30, 1272

\bibitem[\protect\citeauthoryear{Zhdankin, Boldyrev, Perez  \& Tobias}{Zhdankin et~al.}{2014}]{zhdankin_energy_2014}
Zhdankin V.,  Boldyrev S.,  Perez J.~C.,   Tobias S.~M.,  2014, \mn@doi [The Astrophysical Journal] {10.1088/0004-637X/795/2/127}, 795, 127

\bibitem[\protect\citeauthoryear{Zhou, Bhat, Loureiro  \& Uzdensky}{Zhou et~al.}{2019}]{zhou_magnetic_2019}
Zhou M.,  Bhat P.,  Loureiro N.~F.,   Uzdensky D.~A.,  2019, \mn@doi [Physical Review Research] {10.1103/PhysRevResearch.1.012004}, 1, 012004

\bibitem[\protect\citeauthoryear{Zhou, Loureiro  \& Uzdensky}{Zhou et~al.}{2020}]{zhou_multi-scale_2020}
Zhou M.,  Loureiro N.~F.,   Uzdensky D.~A.,  2020, \mn@doi [Journal of Plasma Physics] {10.1017/S0022377820000641}, 86, 535860401

\bibitem[\protect\citeauthoryear{Zhou, Sharma  \& Brandenburg}{Zhou et~al.}{2022}]{zhou_scaling_2022}
Zhou H.,  Sharma R.,   Brandenburg A.,  2022, \mn@doi [Journal of Plasma Physics] {10.1017/S002237782200109X}, 88, 905880602

\bibitem[\protect\citeauthoryear{Zrake}{Zrake}{2014}]{zrake_inverse_2014}
Zrake J.,  2014, \mn@doi [The Astrophysical Journal] {10.1088/2041-8205/794/2/L26}, 794, L26

\bibitem[\protect\citeauthoryear{Zrake}{Zrake}{2016}]{zrake_crab_2016}
Zrake J.,  2016, \mn@doi [The Astrophysical Journal] {10.3847/0004-637X/823/1/39}, 823, 39

\bibitem[\protect\citeauthoryear{Zrake \& Arons}{Zrake \& Arons}{2017}]{zrake_turbulent_2017}
Zrake J.,  Arons J.,  2017, \mn@doi [The Astrophysical Journal] {10.3847/1538-4357/aa826d}, 847, 57

\makeatother
\end{thebibliography}

%%%%%%%%%%%%%%%%%%%%%%%%%%%%%%%%%%%%%%%%%%%%%%%%%%

%%%%%%%%%%%%%%%%% APPENDICES %%%%%%%%%%%%%%%%%%%%%

\appendix

\section{Time evolution and magnetic power spectra}
\label{hel_urms}

\begin{figure}[h!]
    \centering
    %\begin{flushright}
        \includegraphics{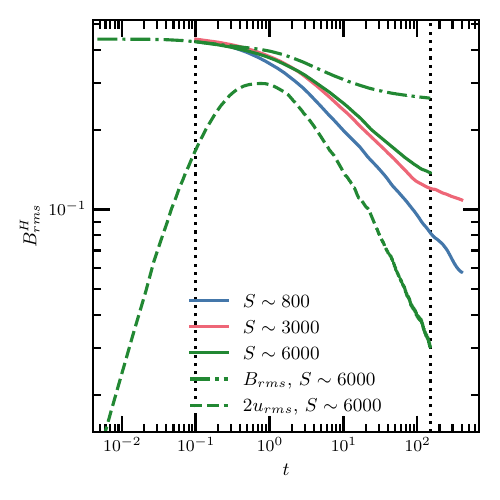}   \includegraphics{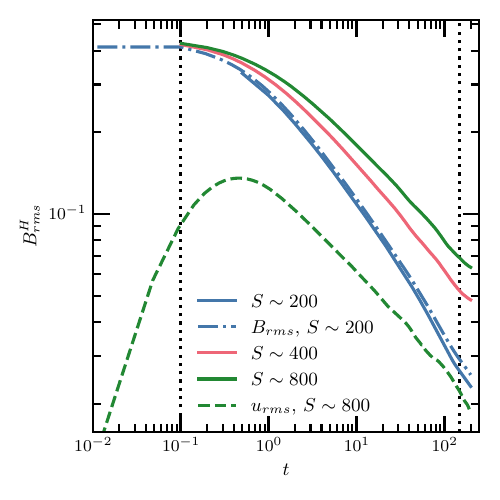} \\
    %\end{flushright}
        \caption{Evolution curves of $B^H_{rms}$ (solid), $B_{rms}$ (dash dotted) and $u_{rms}$ (dashed) with time from fully helical 2.5D (3D) runs on upper panel (lower panel) with different values of $S$. The black dotted vertical lines denote the time range plotted in \Figs{2.5dplots}{3dplots} of Section 3.}   
        \label{urms_plots}
    \end{figure}

The complete evolution of rms values of helical magnetic field, magnetic field and velocity fields with time from helical decaying turbulence at different values of $S$ in 2.5D and 3D are shown in \Fig{urms_plots}.

    \begin{figure}[h!]
    \centering
    %\begin{flushright}
        \includegraphics{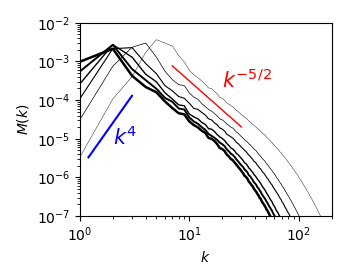}   \includegraphics{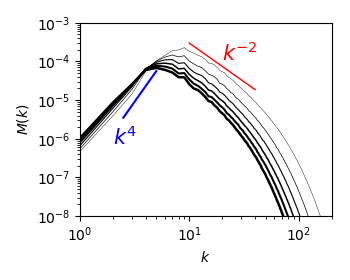} \\
    %\end{flushright}
        \caption{Top panel: Magnetic power spectrum $M(k)$ for 3D helical $1024^3$ run with $S\sim 800$ at different time snapshots $(t = 10, 25, 40, 55, 70)$ with the thickest at the latest time $t = 85$. Bottom panel: Magnetic power spectrum $M(k)$ for 3D nonhelical $1024^3$ run with $S\sim 800$ at different time snapshots $(t = 10, 14, 17, 20, 23)$ with the thickest at the latest time $t = 26$.}   
        \label{ps_hel}
    \end{figure}

\section{Algorithm for calculating Minkowski Functionals}
\label{algoMF}
	In section \ref{localaniso}, we have argued that local anisotropy in a 3D turbulent MHD system can render it quasi-two-dimensional. To characterize this local anisotropy, we analyze the shape of magnetic field isosurfaces. The required algorithm must perform the following tasks,
	\begin{enumerate}
		\item Extract isosurfaces from 3D volumetric data
		\item Calculate geometric quantities of these isosurfaces
	\end{enumerate}
	\subsection{Iso-surface extraction using marching cubes algorithm\label{marching cubes}}
	We use the \textit{marching cubes} algorithm to obtain isosurfaces from the data cube. The input data is a scalar function $f$ on a 3D cubic grid. The algorithm samples through each unit cube, hence the name. The algorithm starts by comparing the value of $f$ at each grid point to the user-provided isosurface value. Then, with the help of a pre-defined look-up table, triangles are placed inside a unit cube, thus constituting a triangulation of the surface. The initial algorithm for the look-up table by \cite{lorensen} had face and internal ambiguities and was later modified by \cite{lewiner} to create a triangulated surface that is topologically consistent. For our work, we use a Python implementation of the \citet{lewiner} algorithm as in \textit{skimage.measure.marching\_cubes} module.
	
	\subsection{Quantifying geometric surfaces using Minkowski Functionals\label{minkowski functionals}}
	
	The \textit{marching cubes} algorithm provides us with the position vector for vertices of the triangles and the gradient direction at each vertex of individual triangles in the order of the flow of the algorithm. To calculate Minkowski Functionals of the triangulated isosurfaces, we closely follow \cite{sheth}. To separate different connected isosurfaces, we identify the triangles that share an edge with a given triangle. 
	
	On a triangulated isosurface, two triangles can share a common edge. Hence, on a connected isosurface, each triangle has three other triangles with which it shares its edges. All the triangles can be viewed together as an undirected graph. Each node of the graph is a triangle. The three edges that come out of a node connect to the the three nodes corresponding to the three triangles that share a common edge with the former. Firstly, we construct a sparse matrix indicating triangles that share common edges. Then, we use a graph algorithm implemented in \textit{scipy.cs\_graph.connected\_components} to identify individual connected structures. To avoid inconsistencies, we impose a lower threshold on the number of triangles making up the isosurfaces. If the number of triangles is less than 10, we reject those isosurfaces.
	\par
\begin{figure}[h!]
		\centering
		\includegraphics[width=0.45\textwidth]{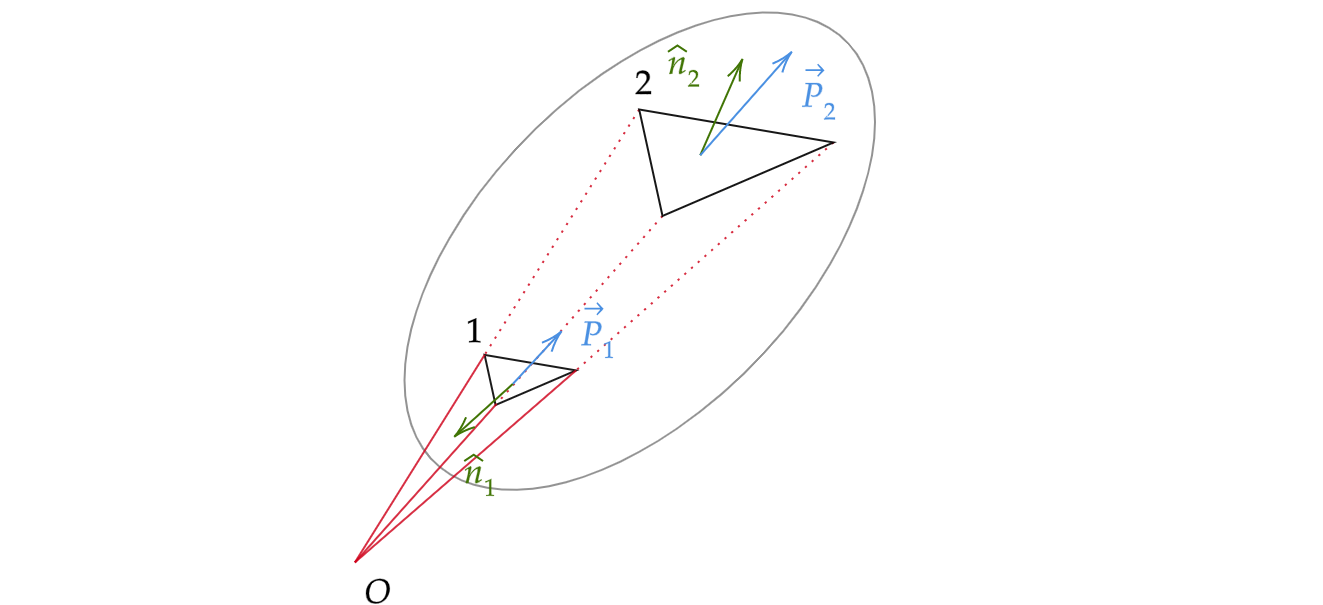}
		\caption{Calculation of volume by summing contributions due to individual tetrahedra formed by the triangles and the origin}
		\label{fig:volume}
	\end{figure}

 We restrict to a particular connected structure and obtain the geometric quantities leading to MFs as follows:
	
	\paragraph*{Surface area}
	The sum of the areas of individual triangles gives the surface area. Using the position vectors for vertices, we calculate the vectors along the edges of the triangles. The area of individual triangles is calculated by taking the absolute value of the cross product of the vectors along any two edges.
	\paragraph*{Volume}
	To calculate the volume, consider one particular triangle on the surface. It subtends a cone at the grid's origin $(0,0,0)$. The volume of the triangular cone is,
\begin{equation}
	V_i = \frac{1}{3}A_i (\bm{P}_c\cdot \nnn),
\end{equation}
where $A$ is the area of the base triangle, $\nnn$ is its normal, and $\bm{P}_c$ is the position vector of the centroid. The height of the cone is given by $\mathbf{P}_c \cdot \nnn$. Now, consider two triangles $1$ and $2$ on the `opposite' sides of the surface as in \Fig{fig:volume}. The triangle $1$'s height is `negative' since the centroid and normal form an obtuse angle. Hence, the volume is `negative'. However, in the case of triangle $2$, the volume is `positive'. It cancels out the negative contribution from triangle $1$, thereby giving us only the volume inside the closed surface. Note that diagonally opposite triangles do not necessarily cancel each other's contribution exactly. We expect that the calculated value converges to the actual value in the limit of a large number of triangles for a given surface. While we described the calculation of the volume subtended at the grid's origin, in practice, we use the centroid of the given structure. However, the above analysis goes through as is.  
	
\begin{figure}[h!]
		\centering
		\includegraphics[width = 0.45\textwidth]{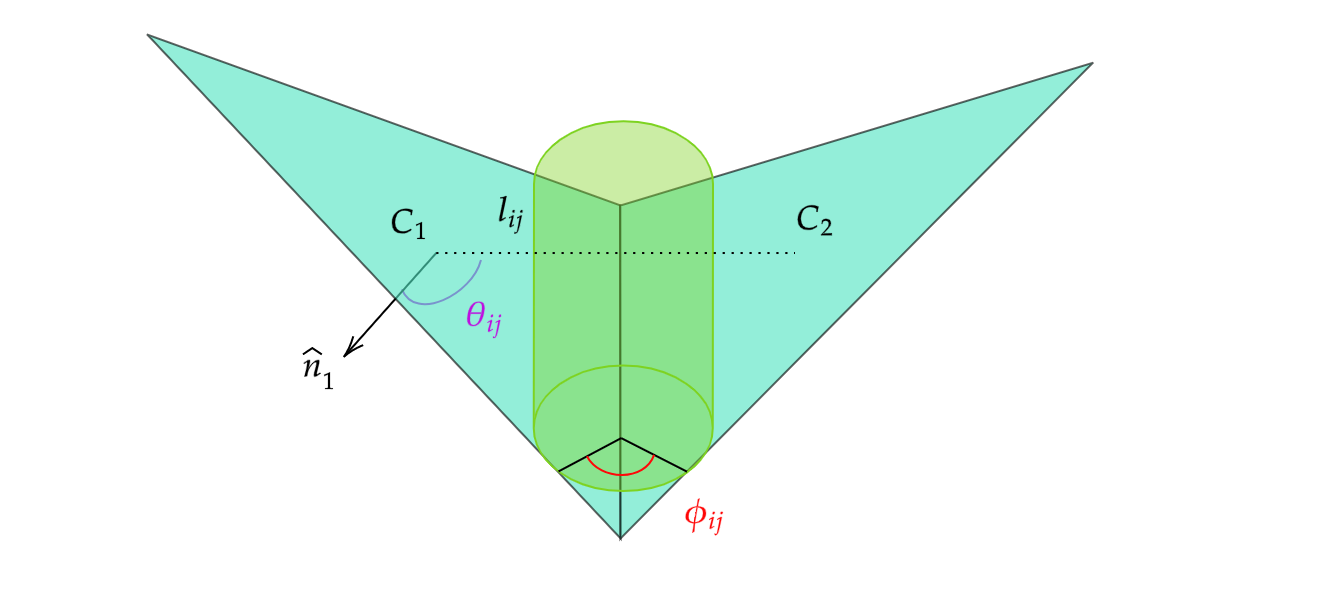}
		\caption{Contribution to IMC from an edge. A cylinder of radius $r$ is fit between the triangular surfaces. They subtend an angle $\phi_{ij}$, and the common edge has length $l_{ij}$. The normal at triangle $1$ and the line joining the triangles $1,2$ subtend an angle $\theta_{ij}$.}
		\label{fig:curvature}
	\end{figure}

	\paragraph*{Integrated Mean Curvature (IMC)}
	The triangulated surface is curved only at the edges of the triangles and not on the faces. To calculate the curvature at the edges, we fit a cylinder of radius $r$ between two adjoining triangles as shown in \Fig{fig:curvature}. The tangents at the point of contact of the two triangles ($i$ and $j$) with the cylinder form an angle $\phi_{ij}$. The area of this curved surface is $\phi_{ij}l_{ij}$, where $l_{ij}$ is the length of the common edge. The principal curvatures of this cylindrical surface is $\kappa_1 = \frac{1}{r}$ and $\kappa_2 = 0$. The two triangles can be oriented caved in(out), contributing negatively(positively) to the IMC. We quantify the orientation $\epsilon$ by calculating the angle between the normal to one of the triangles and the line joining the centroids of the triangles, $\theta_{ij}$ (please see \Tab{orientation}). Hence, the integrated mean curvature of a structure is,
	\begin{multline}
		C = \frac{1}{6\pi}\iint (\kappa_1+\kappa_2)dS = \frac{\epsilon}{6\pi r}\iint dS \\
		= \sum \frac{\epsilon}{6\pi r}rl_{ij}\phi_{ij} =\sum \frac{\epsilon l_{ij}\phi_{ij}}{6\pi}.
	\end{multline}
	\begin{table}[h!]
		\centering
		\begin{tabular}{c | c | c}
			Orientation & $\theta_{ij} $ & $\epsilon$   \\\hline
			Convex & $>\frac{\pi}{2}$& +1\\
			Concave & $<\frac{\pi}{2}$&-1\\
			Flat & $\frac{\pi}{2}$& 0
		\end{tabular}
		\caption{}
		\label{orientation}
	\end{table}
	\paragraph*{Gauss curvature} The Gauss curvature is calculated using the Gauss-Bonnet theorem, which relates it to the Euler characteristic. The Gauss-Bonnet theorem states that
	\begin{equation}
		\label{gauss bonnet}
		\int_M K\,dS = 2\pi \chi(M),
	\end{equation}
	where $K$ = $\kappa_1\kappa_2$. In particular, the above theorem holds also for polyhedral surfaces. Hence, given a triangulation of the surface and thereby the number of vertices $V$, edges $E$, and faces $F$, the Euler characteristic is,
	\begin{equation}
		\chi = V  - E + F = V - \frac{3}{2}F+F = V - \frac{F}{2}.
	\end{equation}
	We have used the fact that each triangle has three edges, two triangles share each edge, and thus, \(E= \frac{3}{2}F\).

\section{Identification of sites of reconnection and nearby magnetic field iso-surfaces }

\label{detailed MF steps}
We compute the MFs for $\vert \BB \vert$ and $\vert \JJ \vert$ fields. While computing MFs for structures throughout the $1024^3$ domain is computationally demanding, it is not necessary as we mainly want to focus on magnetic fields near reconnection sites. 
We identify reconnection sites or grid points which have a large current density value. We quantify this by taking a cutoff value for the current density, $0.7*\vert \JJ \vert_{\text{max}}$. This threshold was chosen to obtain a sufficient number of structures that 
ensure reliable statistics. 

To avoid repetition of processing grid points pertaining to the same current sheet, we isolate a set by the following method. We first apply the threshold (mentioned above) to $\vert \JJ \vert$ to obtain a set of grid-points. Individual clusters of points \textit{belonging} to a current sheet are identified by: \\
(a) Ensuring furthest points in a given cluster are separated at the maximum by $0.5 \ L_{int}$ to prevent the inclusion of neighbouring current sheets, \\
(b) Ensuring that each cluster is separated from another by atleast $1.5 \ L_{int}$. \\
 In this manner, we also avoid double counting of magnetic field isosurfaces. We choose the point with the maximum value among each cluster as the representative point for that cluster. We define the \textit{current sheet} in this cluster to be the isosurface with isovalue $0.5*\vert \JJ \vert_{\text{max, cluster}}$. 
Henceforth, we focus our attention on individual current sheets. For a given current sheet, we restrict our domain size to be $L_{int}$ about its representative point. We do so in order to identify magnetic field isosurfaces that are in proximity to the current sheet.

 To do the MF analysis, we adopt the following scheme to determine, first, the isovalues for $\vert \BB \vert$.  We want to map out the radial profile of $\vert \BB \vert$ around the current sheet. Ideally, an infinite number of rays with all possible orientations are needed, but we use 50 rays with random orientations to get an optimal balance between computational cost and sufficiency.
 We parametrize the rays from $[-0.5,0.5]L_{int}$ with $100$ points. 
 We calculate the $\vert \BB \vert$ value along the rays by linear interpolation.
The maximum and the minimum values of the $\vert \BB \vert$ along each ray nearest to the representative point are obtained. Among these extrema, the largest maximum and the smallest minimum value of $\vert \BB \vert$ are calculated and labelled as $\vert \BB \vert_{\text{max}}$  and $\vert \BB \vert_{\text{min}}$ respectively. Finally, we choose the isovalues for the magnetic field in the following manner,
\begin{multline}\vert \BB \vert_{\text{max}} - \frac{i}{5} \times \left(\frac{\vert \BB \vert_{\text{max}}-\vert \BB \vert_{\text{min}}}{4}\right) \\ \forall i \in {1,2,3,4,5}.\end{multline}
This approach of selecting isovalues is to prevent the isosurfaces from being either minuscule compared to the grid separation or volume-filling. Once the magnetic isosurfaces are identified in this way, MFs can be calculated. Since there can be multiple magnetic field isosurfaces, we want to consider the isosurface closest to the current sheet. So, we calculate the distance between the current sheet and the $\vert \BB \vert$ field isosurfaces\footnote{The distance between two sets, $d(C, D) = min \ d(x_C,x_D)$} and choose the isosurface with the smallest distance to the current sheet.  
 
A particular simulation can be considered one realization from an ensemble of turbulent configurations. The reconnection sites that we obtain form a sample of all possible realizations. Having obtained the MFs of $\vert \BB \vert$ and $\vert \JJ \vert$ isosurfaces, we calculate the sample average and standard error\footnote{It is the standard deviation divided by the square root of the sample size.} of the isosurfaces across all isovalues considered. We then plot the evolution of sample averages with standard errors as error bars.

\section{MF analysis of current sheets}
\subsection{Evolution of $P$ and $F$ for current sheets}
\begin{figure*}
    \includegraphics[width = \columnwidth]{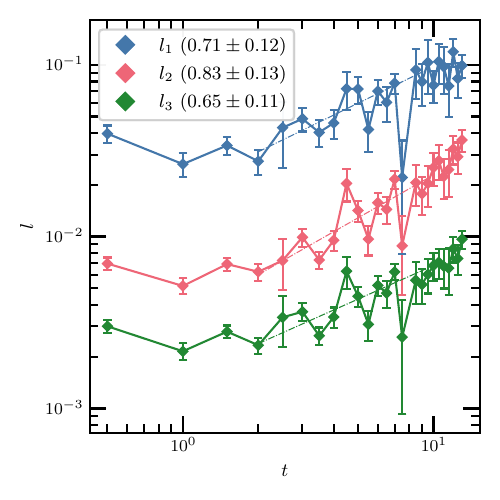}
    \includegraphics[width = \columnwidth]{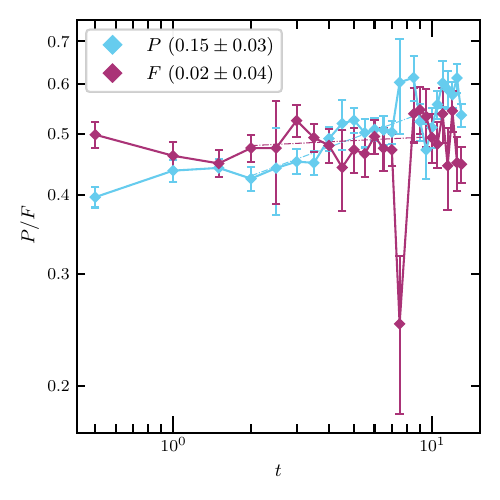}
    \caption{\textbf{Helical:} The time evolution of length scales of \textit{current sheet} isosurfaces is shown in the left panel. The different dashed curves are the least square fit.
    Time evolution of $P$ and $F$ in the early stages of the decay (when the $B_{rms}$ curve behaviour is close to the theoretical prediction of $t^{-2/7}$) is shown in the right panel.
     }
    \label{fig: jj hel evolve}
\end{figure*}

\begin{figure*}
    \includegraphics[width = \columnwidth]{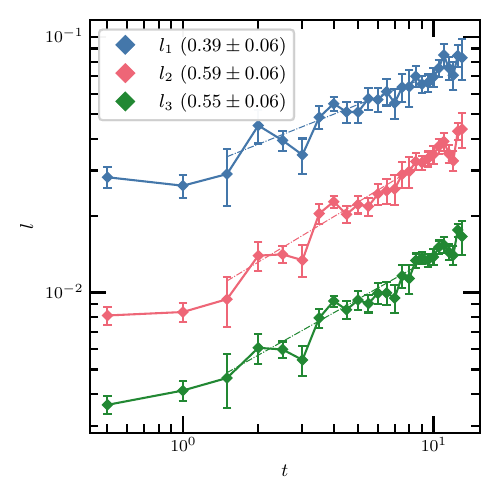}
    \includegraphics[width = \columnwidth]{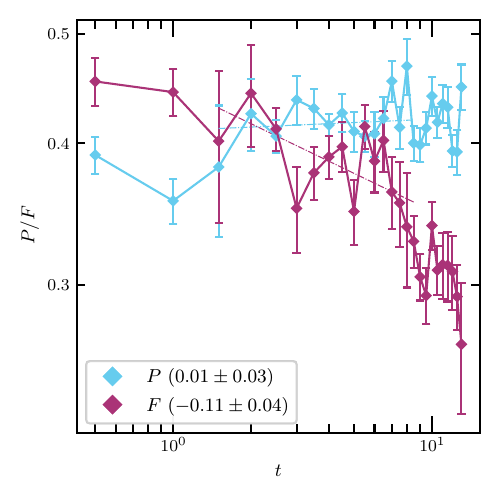}
    \caption{\textbf{Nonhelical:} The time evolution of length scales of \textit{current sheet} isosurfaces is shown in the left panel. The different dashed curves are the least square fit.
    Time evolution of $P$ and $F$ in the early stages of the decay (when the $B_{rms}$ curve behaviour is close to the theoretical prediction of $t^{-1/2}$) is shown in the right panel.
     }
    \label{fig: jj nonhel evolve}
\end{figure*}

Here, we display the evolution of the Minkowski Functionals for the current sheets in \Fig{fig: jj hel evolve} and in \Fig{fig: jj nonhel evolve} for the helical and nonhelical cases respectively. We already noted in \Fig{fig:hel nonhel hydro pf compare}
that the current isosurfaces in both the helical and nonhelical cases are sheet-like.
We plot the evolution of $l_1$, $l_2$ and $l_3$ for the current sheets and observe that they increase with time. This behaviour of the current sheet is consistent with the fact that the system relaxes to larger and larger scales. 
%The scaling of these $l$'s with time needs to be understood. 
We then plot the evolution of $P$ and $F$. The $P$ and $F$ of the current sheets are almost constant with time for both helical $( P \sim t^{0.15 \pm 0.03}, F \sim t^{0.02 \pm 0.04} )$ and nonhelical $( P \sim t^{0.01 \pm 0.03}, F \sim t^{-0.11 \pm 0.04} )$ cases. We also report the linear fits for the length scales $l$ in the plot. 


\section{Results from forced hydrodynamic turbulence}
\label{hydroMF}

\begin{figure*}
    \centering
    \includegraphics[width=0.75\linewidth]{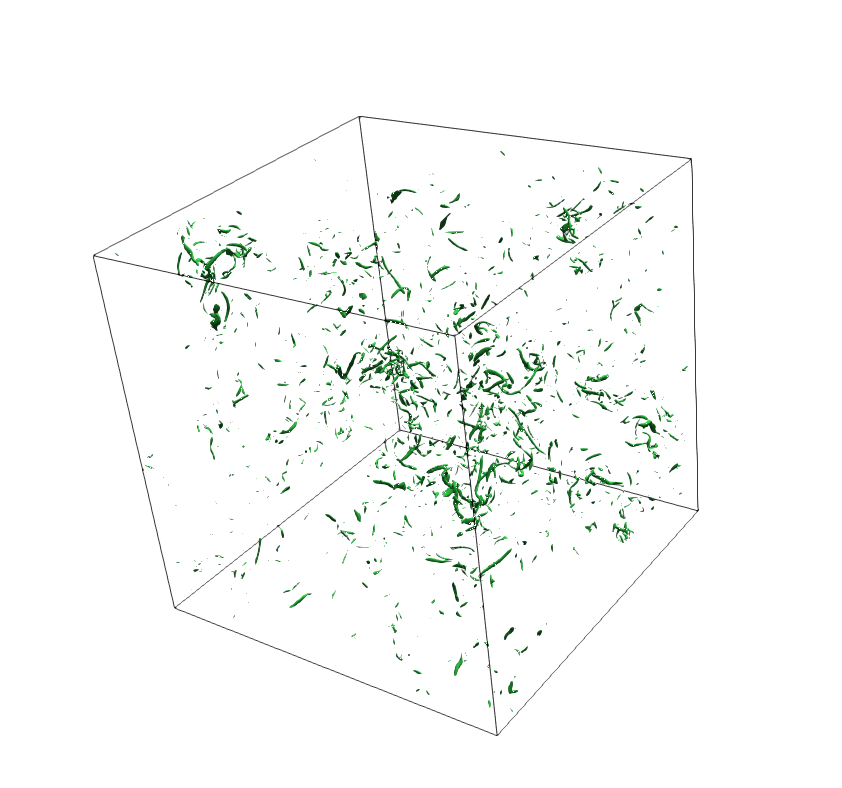}
    \caption{Isosurfaces of vorticity fields at isovalue $5 \bm{\omega}_{\text{rms}}$ of a 
    forced hydrodynamics $512^3$ simulations  at $t=80$. From the figure, it is clear that the 
    structures are akin to elongated tubes.}
    \label{fig:ou iso}
\end{figure*}

\begin{figure*}
    \centering
    \includegraphics[width=0.8\columnwidth]{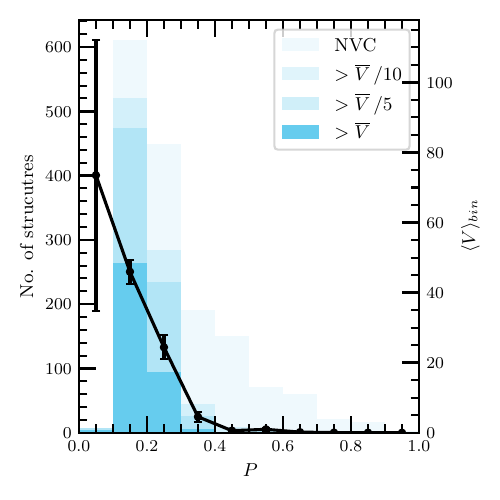}
    \includegraphics[width=0.8\columnwidth]{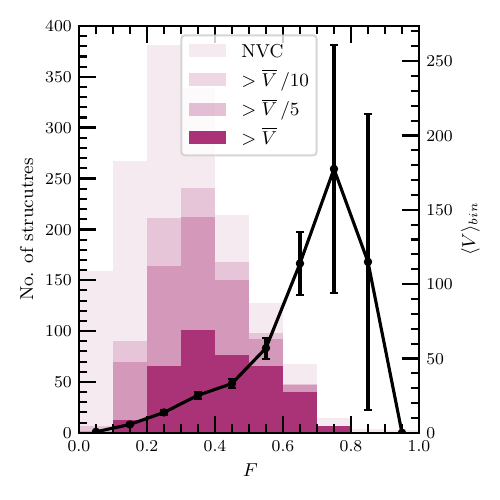}
    
    \caption{Histogram for planarity (left panel) and filamentarity (right panel) of all the vorticity
structures at isovalue $5\bm{\omega}_{\text{rms}}$ in the forced hydrodynamics $512^3$ simulation at $t = 80$. Also shown in the black circles is the average volume in each
bin. NVC stands for no-volume-cutoff. The darker shades represent statistics for structures considered above a certain volume
threshold denoted by $\overline{V}/n$ and $n$ = 1, 5 or 10. The tendency of large structures to have smaller planarity and larger filamentarity
is clear from the graph. The same conclusion can also be reached by progressively considering structures with larger and larger
volumes, as we have shown in this plot.
}
    \label{fig:ou PF pdf}
\end{figure*}

We simulated forced hydrodynamic turbulence with a grid size of $512^3$, average forcing wavenumber $k_f \approx  1.5$, and zero initial velocity field. When the turbulence is fully developed (we confirm the $k^{-5/3}$ spectrum), we perform the morphological analysis using MF. At time $t=80$, we look at the saturation of the velocity field and confirm that turbulence has fully developed in the system.
In 3D hydrodynamic turbulence, the high vorticity isosurfaces are present as tubular structures \citep{hydro_ref}. Hence, we expect structures with isovalues higher than $\bm{\omega}_{rms}$ at a particular time to be statistically filamentary (large $F$). A preliminary visual confirmation is seen in \Fig{fig:ou iso}. We calculate MF of $\bm{\omega}$ at an isovalue of $5 \bm{\omega}_{\text{rms}}$. Initially, we plot the histogram of planarity and filamentarity in \Fig{fig:ou PF pdf}. We plot the mean volume in each bin to understand where the structures of various sizes (volume) contribute. On sequentially removing surfaces with volumes less than a certain fraction of the total mean volume, the peak of the histogram shifts to higher values for filamentarity and lower values for planarity. Hence, larger structures are predominantly filamentarity with an almost circular cross-section (with normal along the long axis). Hence, the system is statistically tube-like. 

% Don't change these lines
\hfill
%\bsp	% typesetting comment
\label{lastpage}
\end{document}